\documentclass[acmsmall]{acmart}

\usepackage{subcaption}
\usepackage{xcolor}
\usepackage{soul}
\usepackage{array}
\AtBeginDocument{%
  }

\begin{document}

% \acmConference
% \acmBooktitle

\setcopyright{cc}
\setcctype{by}
\acmJournal{PACMHCI}
\acmYear{2026} 
\acmVolume{10} 
\acmNumber{2} 
\acmArticle{CSCW026}
\acmMonth{4} 
\acmDOI{10.1145/3788062}

\title[Design Considerations for Supporting Everyday Practices in Community-Based Farming]{"It’s More of a Lifestyle": Design Considerations for Supporting Everyday Practices in Community-Based Farming}

\author{Minghe Lu}
\email{lu000267@umn.edu}
\orcid{0009-0001-0976-3261}
\affiliation{%
  \institution{University of Minnesota}
  \city{Minneapolis}
  \state{MN}
  \country{USA}
}

\author{Zhanming Chen}
\email{chen8475@umn.edu}
\orcid{0000-0002-9913-7239}
\affiliation{%
  \institution{University of Minnesota}
  \city{MN}
  \state{Minnesota}
  \country{USA}
}

\author{May Sunmin Hwang}
\email{smhwang@umn.edu}
\orcid{0000-0001-9552-8606}
\affiliation{%
  \institution{University of Minnesota}
  \city{Minneapolis}
  \state{Minnesota}
  \country{USA}
}

\author{Ji Youn Shin}
\email{shinjy@umn.edu}
\orcid{0000-0003-4978-3897}
\affiliation{%
  \institution{University of Minnesota}
  \city{Minneapolis}
  \state{MN}
  \country{USA}
}

\renewcommand{\shortauthors}{Lu et al.}

\begin{abstract}
    Farming plays a significant role in the economy by supporting related industries such as food, retail, and local services. Community-based small farms, while offering unique social and cultural benefits, face persistent challenges, including limited access to formal education and underdeveloped infrastructure, which have been discussed in prior research. This study focuses on community-driven factors, such as workarounds for recording critical information and practices for passing down farming knowledge across generations. Through 11 semi-structured interviews with farmers from a small ethnic community, the Hmong, we explore how bonding social capital, rooted in close family and community ties, supports informal knowledge exchange and creates pathways to bridging and linking capital. These relationships help farmers connect to broader networks, resources, and institutions. Our findings highlight opportunities for designing technologies that support and strengthen existing support systems. We discuss how technologies should be designed to reflect the cultural values, unique practices, and intergenerational relationships embedded in community-based farms.
\end{abstract}




% Articles V10cscw001-V10cscw043 use
\received{May 2025}
\received[revised]{November 2025}
\received[accepted]{December 2025}


\maketitle

\section{Introduction}
Farming is a critical determinant of the U.S. economy and an essential part of individuals' lives, generating over \$1.5 trillion in output and supporting 22 million jobs~\cite{martin_etal_AgFoodSectors_2025}. The types of farms are highly varied, ranging from industrial farms that dominate large-scale food production, to family farms~\cite{u.s.departmentofagriculture_nationalinstituteoffoodandagriculture_FamilyFarms_2024}, and community-based farms~\cite{bell_FarmingUsAll_2004, u.s.departmentofagriculture_UrbanAgricultureInnovative_}. Among these, community-based farms are locally rooted agricultural operations that center the needs, participation, and well-being of the surrounding community in both food production and farm management~\cite{friedman_etal_UnderstandingFarmersData_2024}. These farms are usually small and have an income less than \$350,000 annually from selling agricultural products, according to the USDA~\cite{farmincometeam_FarmSectorIncome_2025}. They emphasize shared labor or governance, support local food systems, and often incorporate social, educational, or cultural goals. Unlike many industrial farms, they offer unique benefits, including facilitating knowledge exchange among farmers and helping to ease financial pressures and labor shortages~\cite{hayden_etal_ImportanceSocialSupport_2018}.
 
Despite their benefits, they also face challenges, including insufficient infrastructure and limited access to advanced information and technology, which suppresses their potential for growth~\cite{dhillon_moncur_SmallScaleFarmingReview_2023, ito_tanaka_SuccessesChallengesCommunitybased_2020, kuwabara_SociologicalAdvantagesChallenges_2014, rotz_etal_AutomatedPasturesDigital_2019}. For example, many community-based farms lack access to suitable farming tools, making it difficult to track essential information, such as production and seed selection, or connecting with local markets~\cite{christensen_etal_KrishiKonthoAgricultural_2018}. In addition, they lack the technology to cope with unpredictable weather, leading to disrupted planting schedules and lower yields~\cite{idrees_etal_UrduLanguageBased_2019}. Together, these challenges hinder farmers from achieving economic growth through more efficient farming and from creating an ideal work environment that considers their well-being. 

In order to support distinctive challenges of farming communities based on their unique characteristics, HCI and CSCW researchers have suggested various technology-mediated solutions in different contexts~\cite{steup_etal_GrowingTinyPublics_2018, oduor_etal_PracticesTechnologyNeeds_2018, sharma_etal_WhoseParticipationCounts_2024}. They include tools for tracking precise information~\cite{chen_etal_EmpoweringFarmingCommunities_2025} and enhancing farming decision-making~\cite{bi_etal_UnderstandingFarmersExpectations_2023}. For example, Friedman et al. identified limited literacy and knowledge of technology as a key challenge and proposed farm-management-information-systems that can be integrated with their traits and existing information management practices to support it~\cite{friedman2009community}. Another recent study focused on how technologies for tracking and supervising agricultural employees affect farmworkers' well-being, highlighting the need for technologies that promote positive outcomes by supporting their well-being rather than exerting external control and discipline~\cite{doggett_etal_MigrantFarmworkersExperiences_2024}.

Social capital is widely recognized as a critical and inherent resource embedded in social relationships, enabling individuals to access support and facilitating coordination and cooperation for mutual benefit within a group~\cite{coleman_SocialCapitalCreation_1988}. HCI and CSCW researchers have extended the application of this concept across diverse contexts, including farming~\cite{mahaarcha_sirisunhirun_SocialCapitalFarmers_2023}, and disaster resilience, where social capital enables communities to mobilize resources, disseminate information, and support recovery efforts~\cite{aldrich_meyer_SocialCapitalCommunity_2015, makridis_wu_HowSocialCapital_2021}. For example, studies examined the role of social capital in supporting the well-being of individuals and communities through technologies~\cite{dillahunt_FosteringSocialCapital_2014}, and have suggested social capital as an indicator for effectively expanding online networks for optimized information dissemination~\cite{subbian_etal_FindingInfluencersNetworks_2013}. Similarly, social capital has been shown to play a key role in facilitating farmers’ decision-making, knowledge exchange, and participation in rural digitalization efforts~\cite{mathanda_etal_DoesSocialCapital_2025, gao_qiao_HowSocialCapital_2025, kelly_etal_ExploringKnowledgeExchange_2022}.

While existing research has explored the application of social capital in technology design, there is limited understanding of how small-farming communities actively develop and adapt strategies to leverage their existing social capital in addressing external challenges, such as limited access to local retail networks, farming tools, and other critical resources. Given that fully leveraging social capital offers rich opportunities for growth, including accessing grants and collaborating with external groups~\cite{knapik_InnovativeModelCommunitybased_2018}, it is essential to understand the needs and unique characteristics of small-scale farms in order to identify relevant design opportunities. 

This study aims to identify technology design implications that support the values of farming communities and their day-to-day practices. Hmong American farmers show unique traits, including strong social bonds, resourceful adaptation to limited infrastructure, and reliance on both traditional farming skills and collective networks~\cite{neitzel_etal_SafetyHealthHazard_2014}. These characteristics shape how they collaborate and sustain their livelihoods. Working with this small-scale local farming community allowed us to see how social capital and individual strengths can be connected to community-based assets that, in turn, inform technology design opportunities~\cite{yosso_whose_2005,mathie_clients_2003}. To examine these dynamics, we conducted 11 semi-structured interviews with farmers. This study is guided by two research questions:

\begin{itemize}
    \item How do small-scale community farmers manage their farming routines and use social capital to develop strategies in response to environmental and situational challenges? (RQ1)
    \item What design opportunities exist for developing farming technologies that align with the values, practices, and social capital of community-based small farms? (RQ2)
\end{itemize}

By addressing these questions, this paper makes two key contributions to the fields of HCI and CSCW. First, it examines how small-scale community farmers manage their day-to-day routines and adapt to local environmental and situational challenges. Second, it identifies design opportunities for future farming technologies that build on the social capital embedded in community-based small farms, including intergenerational knowledge sharing and strong local relationships, to better align with farmers' lived practices and values. We highlight specific practices that should inform technology design, such as workaround strategies for managing complex farming schedules and leveraging local networks to distribute produce and uphold a sense of responsibility to the local community. 

\section{Related Work}

\subsection{Social Capital in the Design of Sociotechnical Systems}
Social capital, defined as the networks, norms, and trust that enable collective action~\cite{putnam_BowlingAloneCollapse_2001}, has served as a conceptual foundation in research on community development. Researchers have framed social capital from various perspectives, such as a resource embedded in relationships~\cite{bourdieu_FormsCapital_1986} or as a mechanism that facilitates cooperation within closed networks~\cite{coleman_SocialCapitalCreation_1988}. Putnam further distinguishes between bonding capital (e.g., strong ties among individuals with similarities or within close-knit cultural groups) and bridging capital (e.g., weak ties connect across different social groups that provide access to resources)~\cite{putnam_BowlingAloneCollapse_2001}, and Woolcock~\cite{woolcock_SocialCapitalEconomic_1998} introduced linking capital, which refers to connections with individuals in positions of authority or power. This distinction helps clarify how different types of social capital contribute to community resilience.

Social capital is particularly useful for communities in challenging environments, such as natural disasters and public health crises. Previous studies in sociology, economics, and HCI have explored its critical role in helping social groups and communities cope with and recover from challenges~\cite{chamlee-wright_storr_SocialCapitalCollective_2011, woolcock_narayan_SocialCapitalImplications_2000, hsiao_dillahunt_TechnologySupportImmigrant_2018}. For example, after the 2011 tsunami in Japan, communities with strong pre-existing social ties demonstrated faster and more robust post-disaster recovery, while more isolated communities struggled ~\cite{aldrich_meyer_SocialCapitalCommunity_2015}. Similar patterns emerged during the COVID-19 pandemic in the U.S., where regions with higher levels of social capital (e.g., family unity) experienced significantly lower infection and death rates~\cite{makridis_wu_HowSocialCapital_2021}. Other studies have shown that social capital contributes to various forms of resilience, including improved regional economic performance~\cite{knack_keefer_DoesSocialCapital_1997}, increased farming yields~\cite{sutherland_burton_GoodFarmersGood_2011, krishna_ActiveSocialCapital_2002}, and improved mental health in low-income areas~\cite{ledogar_fleming_SocialCapitalResilience_2008, nair_etal_2299SocialCapital_2013}, underscoring the importance of emotional safety and mutual aid. These examples illustrate how social capital can supplement more tangible resources by enhancing disaster recovery, reducing public health risks, supporting economic productivity, and fostering mental well-being in underserved communities.

Technology emerges as a potential facilitator of social capital. Information and communication technologies (ICTs) are powerful tools for fostering social capital across diverse groups~\cite{burke_etal_SocialCapitalFacebook_2011, kameswaran_etal_SupportSocialCultural_2018, dillahunt_FosteringSocialCapital_2014}. Previous literature has explored the effectiveness of ICTs in supporting individuals from marginalized communities, including those with limited education, and those from low-income and low-resource communities, in developing both bonding and bridging capital~\cite{dillahunt_FosteringSocialCapital_2014, hui_etal_HighRiskHigh_2023, anuyah_etal_CharacterizingTechnologyNeeds_2023}.

Social platforms are one common form of technology, and are widely recognized for their ability to connect people with resources that help them pursue various goals. Studies have shown that social media use can facilitate emotional support. For example, platforms like Facebook help users who are actively seeking interactions optimize the value of their networks by increasing their sense of connection~\cite{burke_etal_SocialCapitalFacebook_2011}. Additionally, these platforms expand users’ networks beyond their immediate social circles, connecting them to broader ranges of communities and enhancing access to emotional support~\cite{cornejo_etal_AmbientAwarenessStrengthen_2013, lei_etal_UnpackingICTsupportedSocial_2024}.

Social capital has also been actively studied in family settings to understand how unique dynamics within close networks, such as intergenerational interactions and collaboration, can be supported through technology. These approaches offer insights into how social capital is cultivated and sustained, and for what purposes. Even though studies do not necessarily point directly to the concept of social capital, they explore intergenerational cooperation and interaction, which are effective ways to promote social capital~\cite{chen_i_2025}. For example, research has shown that engaging individuals from different generations (e.g., younger and older adults) in collaborative games (e.g., virtual reality escape rooms) can reduce ageism among younger participants~\cite{zou_etal_MitigatingAgeismVirtual_2024}. Similarly, studies have found that interventions designed for grandparents and grandchildren to use together can strengthen their connection and enhance intergenerational interactions~\cite{xu_etal_CoRootCollaborativePlanting_2024, butzer_etal_GrandtotemSupportingInternational_2020}. Collectively, these studies provide diverse lenses on social capital and show how it can guide technology design to yield positive outcomes, including stronger relationships.

\subsection{Agricultural Technologies to Support Farming Communities}
Advanced technologies, such as the Internet of Things (IoT), autonomous robotics, and artificial intelligence (AI), have created promising opportunities to enhance the precision and efficiency of farming~\cite{abbasi_etal_DigitizationAgriculturalIndustry_2022}. These technological solutions were developed to support climate-adaptive farming~\cite{adenle_etal_GlobalAssessmentTechnological_2015}, livestock management~\cite{neethirajan_etal_RecentAdvancementBiosensors_2017}, and food safety surveillance~\cite{griesche_baeumner_BiosensorsSupportSustainable_2020}. For example, a system that integrates sensor data and machine learning algorithms was designed to provide real-time recommendations for crop management, such as fertilizer use~\cite{islam_etal_MachineLearningEnabled_2023}.

Despite the effectiveness of these systems, existing technologies primarily focus on the needs of large-scale, commercially-oriented farms in the Global North~\cite{doggett_etal_HCIResearchAgriculture_2023}. To support a broader range of farms, HCI and CSCW studies have identified key challenges faced by small-scale farmers, including limited infrastructure and connectivity, restricted access to weather and market information, and other essential resources, as well as low digital-literacy levels~\cite{doggett_etal_HCIResearchAgriculture_2023, heldreth_etal_WhatDoesAI_2021}. 

Various technologies have been introduced to help small-scale farmers access agricultural information and resources more easily~\cite{singh_etal_WhatsPlacePlatformization_2025,wyche_steinfield_WhyDontFarmers_2016, yang_etal_DetectingFunctionalField_2020}. For example, one study investigated satellite-driven field mapping aimed at assisting small-scale farmers in developing countries in utilizing their farmland and allocating resources more efficiently~\cite{yang_etal_DetectingFunctionalField_2020}. Similarly, mobile platforms tailored for farmers can offer access to agricultural advice, market price alerts, and weather updates, thereby enabling better-informed decision-making~\cite{wyche_steinfield_WhyDontFarmers_2016}. Another group of studies examined approaches to address farmers' low digital-literacy~\cite{friedman_etal_UnderstandingFarmersData_2024,raghunath_etal_EKichabiV2Designing_2024}. For instance, Friedman et al. suggested digital tools that require minimal data entry and adapt according to users' literacy levels to enhance informed farming practices~\cite{friedman_etal_UnderstandingFarmersData_2024}, and others emphasized the role of Android-based applications in improving access to agricultural information among rural populations with limited connectivity~\cite{raghunath_etal_EKichabiV2Designing_2024}.

\subsection{Information Tracking in Different Social Contexts}

Information tracking is a crucial part of daily life, as it allows individuals to organize, retrieve, and use data effectively to achieve various goals across different contexts~\cite{ciolfi_lockley_WorkLifeBack_2018, bressa_etal_DataEveryDay_2022}. In addition to personal settings, information tracking offers significant benefits in professional and industrial domains. For example, it facilitates improved decision-making in farming~\cite{chen_etal_EmpoweringFarmingCommunities_2025}, exhibits the potential of enhancing workplace productivity~\cite{kim_etal_UnderstandingPersonalProductivity_2019}, and enables more accurate performance monitoring in various social contexts, including healthcare and field work~\cite{bakewell_etal_EverythingWeEverything_2018, okane_etal_TurningPeersIntegrating_2016}. Researchers have shown that information tracking can improve learning outcomes~\cite{rong_etal_UnderstandingPersonalData_2023}, help patients gain a sense of control during chronic disease recovery~\cite{ayobi_etal_QuantifyingBodyCaring_2017}, and enhance personal well-being by connecting data to emotionally meaningful milestones, such as completing marathons, reaching personal fitness goals, or raising funds for charity~\cite{rooksby_etal_PersonalTrackingLived_2014}.

As digital technologies become increasingly embedded in both everyday and work-related environments, understanding the factors that facilitate or hinder information tracking practices has become critical for designing context-specific, tailored tools in HCI~\cite{bergman_etal_PersonalInformationManagement_2004, rooksby_etal_PersonalTrackingLived_2014}. Epstein et al. discussed the essence of designing effective information tracking strategies, including adaptive strategies based on individuals’ needs at different stages, communication with authorities about the data collected, and secured tracking environment that avoids data disclosure~\cite{epstein_etal_ExaminingMenstrualTracking_2017}.

While earlier studies explored the efficiency and effectiveness of personal information tracking, more recent research has examined the role of social dynamics and collaboration in capturing and managing information. For example, using digital tools collaboratively for various healthcare purposes, including mental health tracking~\cite{murnane_etal_PersonalInformaticsInterpersonal_2018}, domestic health management~\cite{shin_etal_MoreBedtimeBedroom_2022}, and patient-caregiver partnerships~\cite{mishra_etal_SupportingCollaborativeHealth_2018, min_etal_UnderstandingContextsChallenges_2023}, can support everyday and long-term health management. Other studies have suggested that digital tools can enhance information tracking in learning contexts~\cite{biedermann_etal_UseDigitalSelfcontrol_2024}, work settings~\cite{moller_etal_CanWorkplaceTracking_2021}, and community-focused initiatives~\cite{saha_etal_CommunityVoiceData_2023}. Moreover, as one of the most prominent technological trends in recent years, generative AI tools are considered promising in supporting information tracking. Anuyah highlighted both the challenges and opportunities of using AI tools, especially large language models, to improve knowledge management within community social service organizations~\cite{anuyah_etal_CharacterizingTechnologyNeeds_2023}.

Several studies have explored the benefits of an effective information tracking system in low-resource communities. For example, Cruz discussed how making computing technologies more accessible can support health information tracking in low-income populations~\cite{cruz_etal_EquityWareCodesigningWearables_2023}. Similarly, Nakikj studied online health communities~\cite{nakikj_mamykina_LostMigrationInformation_2018}, which often lack financial and technical resources, and found that issues like information overload can be addressed through well-designed information tracking strategies. Other studies have shown that mobile technologies can help address information tracking challenges in specific settings, such as animal shelter volunteer communities with limited resources~\cite{zhu_etal_FETCHFosteringEnhancing_2024} and rural farming communities~\cite{saha_etal_CommunityVoiceData_2023}.

Although recent HCI and CSCW work has highlighted social capital as a promising lens for supporting community strengths, small-scale farming communities remain underrepresented, especially those whose members rarely have the opportunity to voice their needs or priorities in academic research. While existing studies have explored the role of social capital in technology design, there is limited understanding of how small farming communities actively develop and adapt strategies to leverage their social capital in addressing external challenges, including limited access to local retail networks, farming tools, and other essential resources. In addition, although agricultural technologies have advanced in areas, such as precise information tracking to improve efficiency and productivity, the unique conditions of small-scale, community-based farms require different information management practices that do not align with those used in large-scale operations. Building on prior work in community social capital, agricultural technologies, and information tracking, our study identifies technology design opportunities tailored to the needs of small-scale farming communities.

\section{Methods}

\subsection{Study Context: Small Community-Based Farming}
This study involved farmers from a small ethnic group called the Hmong, who have a unique background. The Hmong, an ethnic group from Southeast Asia, resettled in large numbers in the U.S.. By 2023, the Hmong population in the U.S. was estimated at approximately 360,000~\cite{gerdner_HealthHealthCare_2024}. As a resettled community, many have continued farming by drawing on traditional agricultural knowledge, using it as a primary means of livelihood and daily food supply in their new environment.

After resettlement, Hmong farmers have contributed to small-scale farming communities by preserving traditional agricultural knowledge, diversifying local food systems, and supporting sustainable practices through crop variety and natural methods. Their experiential farming knowledge includes techniques such as intercropping (planting multiple crops together to improve soil and yield), natural pest control using herbs, seed saving, and crop rotation to maintain soil health. They also strengthen local economies and community networks by actively participating in farmers markets, mentoring emerging farmers, and advocating for communities in similar situations, such as other resettled refugee groups.

Despite rebuilding their lives in a new region, potential challenges related to economic instability and the long-term sustainability of small-scale farming remain significant obstacles for farmers~\cite{neitzel_etal_SafetyHealthHazard_2014, pathak_systematic_2019}. These issues stem from differences between their original and current environments, including cultural factors (e.g., language barriers)~\cite{doggett_etal_MigrantFarmworkersExperiences_2024,liu_eaton_BarriersTrainingHmong_2023} and environmental and structural factors (e.g., different climate conditions, more regulated markets, limited access to farming infrastructure) in the U.S.~\cite{carlisle_securing_2019}. Key challenges include the lack of suitable tools for both direct farming activities (e.g., planting and harvesting) and supporting tasks (e.g., information tracking and management)~\cite{marshall_australian_2020}. These limitations lead to missed opportunities, including restricted land access that is often limited to short-term lease arrangements~\cite{neitzel_etal_SafetyHealthHazard_2014}, as well as minimal adoption of advanced agricultural technologies. As a result, many farmers must rely on less effective methods for tracking yields, managing finances, applying for grants, and optimizing their farm operations.

Studying the practices of Hmong farmers is critical, as their creative workarounds and reliance on social capital demonstrate how individual capabilities can be transformed into collective assets, offering insights that can inform interventions for other small-scale, community-based farming groups.

\subsection{Data Collection}

The goal of this study was to explore the experiences of small community farmers and identify opportunities for technology-mediated smart farming tools. To achieve this, we engaged directly with participants and created opportunities for them to reflect on their lived experiences related to farming. We began with preliminary observations, followed by semi-structured interviews. These included multiple site visits to local organizations that support farmers, farmers’ markets where Hmong farmers sell their produce, and farmlands where we observed their everyday practices. The purpose of these activities was to build background knowledge in farming, establish rapport with the community, and develop interview questions suited to farmers with unique cultural and linguistic backgrounds, rather than to serve as direct data collection for the study.
This groundwork allowed us to construct effective data collection materials, connect more genuinely with our 11 participants, and approach each interview as a meaningful opportunity to explore community-based practices within their specific context.

\begin{figure*}[h]
  \centering
  \begin{subfigure}[t]{0.48\textwidth}
    \centering
    \includegraphics[width=\linewidth]{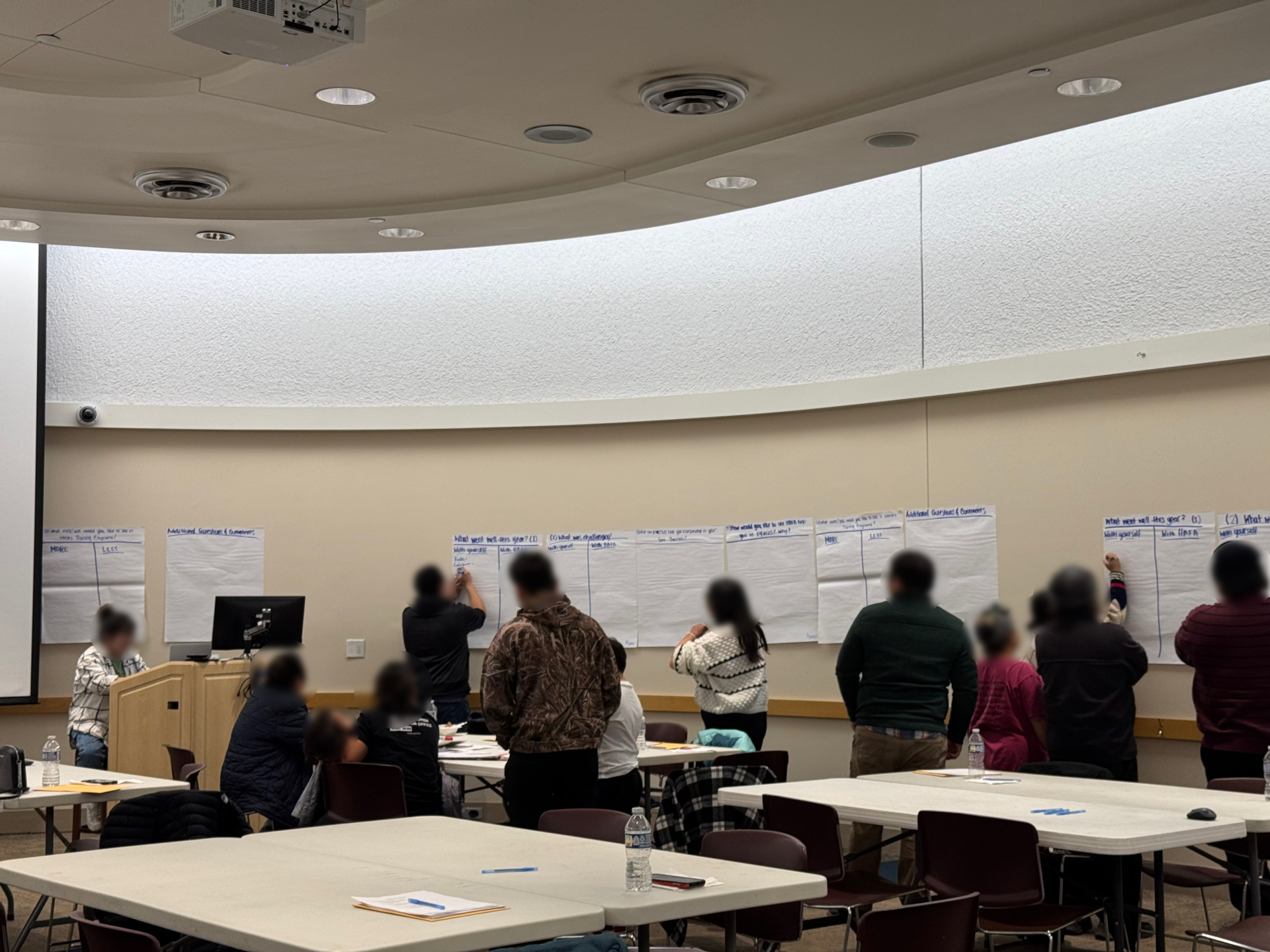}
    \caption{As part of our preliminary observations, we attended a reflection meeting hosted by a local agency.}
    \label{fig:train}
    \Description{As part of our preliminary observations, we attended a reflection meeting hosted by a local agency.}
  \end{subfigure}
  \hfill
  \begin{subfigure}[t]{0.48\textwidth}
    \centering
    \includegraphics[width=\linewidth]{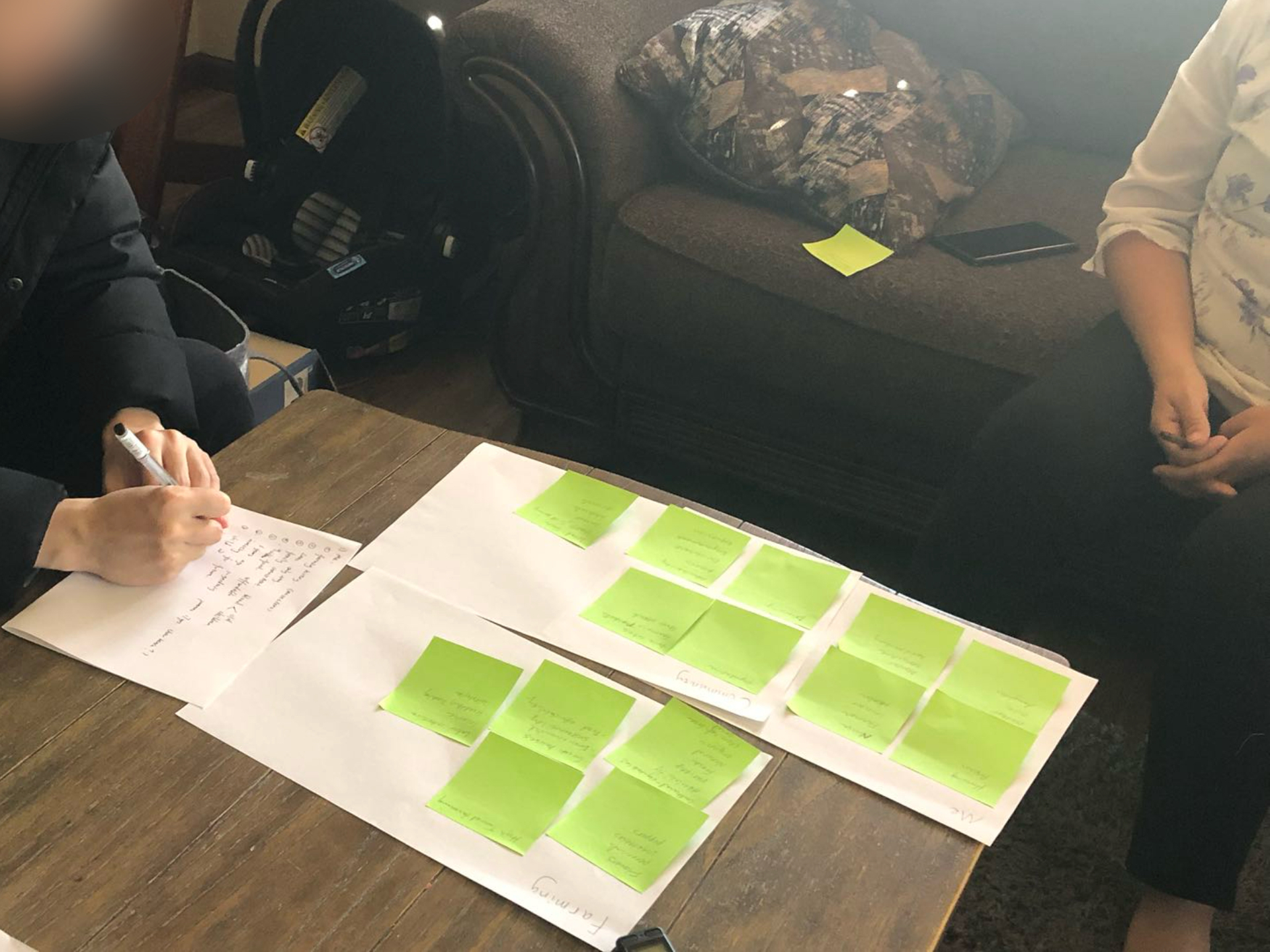}
    \caption{During the mapping activity, a participant explained the meaning of the keywords.}
    \label{fig:map}
    \Description{During the mapping activity, a participant explained the meaning of the keywords.}
  \end{subfigure}
  \caption{Observation and mapping activities.}
  \label{fig:combined1}
\end{figure*}

%%\begin{figure}[h]
%%  \centering
%%  \includegraphics[width=0.6\linewidth]{Figure-Training blur.jpg}  
%%  \caption{During one observation session, a staff member conducted a training session for Hmong farmers that included selecting packaging boxes.}
%%  \label{fig:train}
%%  \Description{During one observation session, a staff member conducted a training session for Hmong farmers that included selecting packaging boxes.}
%%\end{figure}

\subsubsection{Preliminary Observations and Rapport Building} \hfill

\noindent Our participants are Hmong farmers who form ethnic enclaves in the Minneapolis–Saint Paul metropolitan area of the U.S. While many second- and third-generation community members are more easily integrated into mainstream society, older generations (e.g., individuals over 60) often maintain traditional lifestyles, primarily speak the Hmong language, and interact mostly within their own communities. Due to these unique characteristics, gaining first-hand insights from these farmers was essential to our study, but also posed challenges in building trust and accessing the community.

To form relationships with local farmers, we visited two nonprofit organizations that support farmers in the area. These organizations serve as community hubs where farmers can access information, participate in educational sessions, and lease land and tools.

From spring to winter of 2024, both organizations allowed our team to conduct on-site observations and take part in several farming activities. This helped us better understand the farming conditions of small, community-based farms, the seasonal tasks farmers manage, and, most importantly, become familiar with the community. For example, we attended educational sessions where one organization provided guidance on produce packaging and crop sorting, as well as reflection meetings where farmers shared their experiences and discussed their farming practices (see Figure~\ref{fig:train}). We also visited two different farmers' markets where Hmong farmers sell their produce to observe the types of crops they offer (e.g., lemongrass, mustard greens, Thai chilies) and how they interact with local consumers. Because our visits extended over several months, we were able to observe various farming routines, including planting seeds, harvesting, selling crops, and preparing for overwintering.

These visits prior to recruitment enabled us to build trust with farmers in the community and deepened our understanding of their general farming practices and cultural context. The relationships we formed also shaped our recruitment process, the design of interview questions, and our overall approach as researchers.

\subsubsection{Semi-structured Interviews} \hfill

\noindent We conducted semi-structured interviews with participants who are members of the Hmong farming community. Staff members from the organizations asked whether farmers were interested in participating, and if so, our team visited the farmers at their homes or farmlands, depending on their preferences. One interview was conducted online based on participant availability and convenience. Each interview, lasting 45 to 60 minutes, was conducted by a pair of researchers who supported the process by asking follow-up questions and handling logistics, such as setting up voice recorders and taking notes. For 5 participants whose primary language was not English, they were welcome to bring a family member, friend, or organization staff member to assist with translation, and all of them chose to do so. The semi-structured interviews consisted of two parts: a mapping activity (serving as an icebreaker) and a follow-up discussion (the main data collection), which was audio-recorded for transcription and analysis.

%%\begin{figure}[h]
%%  \centering
%%  \includegraphics[width=0.6\linewidth]{Figure-Mapping.jpg}  
%%  \caption{A participant was generating keywords during the mapping activity, with an organization staff member serving as interpreter.}
%%  \label{fig:map}
%%  \Description{A participant was generating keywords during the mapping activity, with an organization staff member serving as interpreter.}
%%\end{figure}

The first part focused on a mapping activity designed to give participants space to share information, stories, and reflections, allowing them to actively guide the conversation. As an effective icebreaker, this allowed participants to express their thoughts freely, without being limited by how researchers define farming or describe it~\cite{clarke2003situational, lee2017collaborative}. Participants were given three keywords—Me, Farming, and Community—along with sticky notes to generate 15 to 20 words, phrases, or short sentences (see Figure~\ref{fig:map}). They placed notes under the relevant keywords to indicate connections or between keywords to show overlapping relevance. Notes could also be grouped together to show relationships among concepts. Once participants indicated they had no further notes to add, they were invited to explain the meaning or story behind each note. They were also encouraged to continue contributing new thoughts as they spoke. Example keywords and phrases generated by the participants include “variety,” “resources,” and “always thinking about how to make things better each year.” The conversations about the map were audio-recorded and analyzed as part of the interview data. The keywords and completed mapping activities will be reported in a separate manuscript. This mapping activity helped us capture genuine, unfiltered insights and provided a foundational understanding of each participant’s background and farming experience, which informed the second part of the interview.

The second part of the semi-structured interview focused on a set of open-ended questions designed to explore participants' farming experiences in depth. These questions were informed by our preliminary observations and by previous studies on farming in low-resource communities. We used initial guiding questions as prompts to encourage storytelling and reflection. Participants were invited to share their experiences related to farming practices in Hmong culture, seasonal routines, and their experiences farming in the U.S. We also asked about tools or technologies used in farming. These questions were adapted to each participant’s mapping outcomes. For instance, if a participant generated keywords “children” during their mapping activities, we naturally asked more questions about how their children perceived farming or their engagement in it. If a participant had a keyword, “tool,” we asked follow-up questions to further clarify whether they meant hand tools or digital tools in their farming practices.

\subsubsection{Ethical Considerations} \hfill

\noindent Given the characteristics of our participants, many of whom have historically been excluded from traditional academic research, throughout the study, we made it a priority to create a safe and comfortable space where they would not feel tension or discomfort. Participants were given the option to choose the location of the interview, and we approached each session as a conversation rather than a formal data collection procedure. By carefully positioning ourselves as listeners rather than as experts with greater knowledge or authority, we encouraged participants to speak freely and expand on any topic that emerged during the conversation, often leading to unexpected and valuable insights. For example, one participant began by describing his years in farming but gradually reflected on his life trajectory, including his teenage years in France. Throughout the interviews, we supported open sharing through active listening, follow-up questions, and a flexible interview structure. Responses often addressed multiple themes beyond the initial prompt, and we moved forward only when participants had no additional thoughts to share.

\subsection{Participants}
To be eligible, participants had to meet the following criteria: (1) be of Hmong origin, (2) be over 18 years old, (3) have engaged in farming for more than one year, and (4) be willing to participate in the interview in either English or Hmong. 

Through two local farming organizations, we recruited 11 farmers who agreed to participate in the interviews (see Table~\ref{tab:demog}). Among these participants, 8 out of 11 have more than 10 years of farming experience, and 5 out of 10 have a level of education lower than a high school degree. The income level of participants varied from less than \$10,000 to \$99,999, while most participants (8 out of 11) have an income less than \$49,999, which is lower than or close to the local low-income threshold.

We stopped recruitment after the 11th participant, as we had reached theoretical saturation, meaning no new properties of our analytic categories were emerging. Research with focused groups often reaches saturation more quickly when the research question is narrow and the population is relatively homogeneous~\cite{hennink_sample_2022}. By this point, we observed no new information or ideas emerging, which aligns with established criteria for data saturation~\cite{charmaz_ConstructingGroundedTheory_2014}.

\begin{table*}
  \caption{Demographic Information of Participants}
  \label{tab:demog}
  %%\begin{tabular}{lllllll}
  \setlength\tabcolsep{2pt}
  \begin{tabular*}{\linewidth}{@{\extracolsep{\fill}} llllllll }
    \toprule
    %%TERRAIN\footnote{This is a table footnote. This is a table footnote. This is a table footnote.} & (200\,m$\times$200\,m) Square\\
    Participants & Gender & Age & Income & Education \\
    \midrule

    P1 & M & 48 & \$35,000--\$49,999 & High school graduate or GED \\
    P2 & F & 41 & \$35,000--\$49,999 & Some college or 2-year degree \\
    P3 & M & 41 & \$50,000--\$74,999 & Some college or 2-year degree \\
    P4 & F & 37 & \$75,000--\$99,999 & 4-year college graduate \\
    P5 & M & 49 & \$35,000--\$49,999 & Some college or 2-year degree \\
    P6 & F & 35 & \$35,000--\$49,999 & 4-year college graduate \\
    P7 & F & 63 & \$25,000--\$39,999 & Lower than elementary school \\
    P8 & M & 64 & \$25,000--\$39,999 & Lower than elementary school \\
    P9 & F & 61 & \$25,000--\$39,999 & Lower than elementary school \\
    P10 & M & 71 & Less than \$10,000 & Elementary school \\
    P11 & F & 63 & Not disclosed & Lower than elementary school \\

  \bottomrule
  
\end{tabular*}
\bigskip
%%\footnotesize\emph{*} F=Female, M=Male, N/A=Not available.
\emph{Note:} F=Female, M=Male.
\end{table*}

\subsection{Data Analysis}
Audio recordings of the interviews were transcribed for data analysis. We conducted thematic analysis~\cite{braun_clarke_ThematicAnalysis_2012} to identify emerging themes regarding the farming experience of Hmong farmers from community-based small farms. We entered the collected data into NVivo and conducted a four-stage coding process, which involved familiarizing ourselves with the transcripts, performing descriptive and open coding, aggregating codes, and creating affinity diagrams. Throughout the process, we carefully read each transcript and revisited the original audio recordings whenever clarification was needed.
We particularly focused on recognizing the unique characteristics of community-based small farms and farmers’ experiences and routines related to these characteristics. We paid primary attention to the personal stories and their experiences with the community and other community members, which aims to understand the internal factors that impact their farming experiences. To do so, we specifically focused on: (1) experiences with existing farming technologies; (2) perspectives on the characteristics of their farm and farming practices; (3) experiences before, during, and after they joined the farming community; and (4) dynamics among community members.

Three authors participated in the data analysis and met regularly to share progress, discuss unclear codes, and identify emerging themes. We moved to the next phase only after reaching an agreement on the results. The analysis followed multiple stages, each resulting in a corresponding level of code (see Table~\ref{tab:Code}): 602 low-level codes (e.g., “Farmers want to provide the community with fresh vegetables and fruits and help community members stay healthy.”), 127 mid-level codes (e.g., “Mismatches between available farming equipment and tools and farmers' specific use cases.”), and high-level codes (e.g., “Adapting planning and tracking practices for diverse and rotational farming needs.”). In total, we identified four main themes across the data, which we present in the Results section.

\begin{table*}
\caption{Coding Hierarchy and Example Codes}
\label{tab:Code}

\footnotesize

\begin{tabular}{m{0.18\textwidth} p{0.78\textwidth}}
\hline
\multicolumn{1}{c}{Level of Codes} &
\multicolumn{1}{c}{Code Examples} \\ \hline

\centering{Low-Level} &
\begin{minipage}[h]{\linewidth}
\vspace{4pt}
\begin{itemize}
    \setlength\itemsep{1pt}     
    \item Farmers think each crop has different growing and planting times, and these timings affect whether they can achieve a good harvest.
    \item Tractors can't cover the whole land precisely, and farmers still need to use hand tools for details.
    \item Farmers want to provide the community with fresh vegetables and fruits and help community members stay healthy.
    \item Farmers believe that older generations are open to having younger generations participate in farming.
\end{itemize} 
\vspace{4pt}
\end{minipage}\\ \hline

\centering{Mid-Level} &
\begin{minipage}[h]{\linewidth}
\vspace{4pt}
\begin{itemize}
    \setlength\itemsep{1pt}
    \item Farmers' desire for precise and accurate farming plans, including info on crop rows, planting times, and harvest schedules.
    \item Mismatches between available farming equipment and tools and farmers' specific use cases.
    \item Farmers prioritizing the wellbeing and health of community members.
    \item Farmers' preservation of traditional farming knowledge by passing it from older to younger generations.
\end{itemize} 
\vspace{4pt}
\end{minipage}\\ \hline

\centering{High-Level} &
\begin{minipage}[h]{\linewidth}
\vspace{4pt}
\begin{itemize}
    \setlength\itemsep{1pt}
    \item Adapting planning and tracking practices for diverse and rotational farming needs.
    \item Addressing equipment gaps across varied small-scale farming contexts.
    \item Strengthening community-based economic, emotional, and knowledge support.
    \item Fostering intergenerational collaboration to sustain future farming pathways.
\end{itemize} 
\vspace{4pt}
\end{minipage}\\ \hline

\end{tabular}

\end{table*}

\section{Results}
During the interviews, participants reflected on the meaning of farming in their lives. They shared that they had learned hands-on farming skills from the older generation and emphasized that farming has long been a tradition in Hmong culture, where nearly every family participated. For them, farming became an essential routine they wished to preserve while living in a new environment. In this section, we present key themes that emerged from their experiences: the unique approaches to planning and tracking shaped by the characteristics of small community farms; the need for tools adaptable to diverse and often irregular farm plots; the strength of community-based resources; and the desire to sustain farming across generations by creating opportunities for younger members of the community.

\subsection{Adapting Planning and Tracking to Fit Diverse and Rotational Farming Practices}
We found that Hmong farmers cultivate a large variety of crops as a way to respond to uncertainty, which is shaped in part by low-resource conditions, such as inclement weather and limited access to affordable farmland. This practice, in turn, requires more time, involves land rotation, and accounts for the short growing cycles of certain crops. In this context, information tracking becomes particularly important, as even small issues such as forgetting which plots were fertilized, where certain crops were planted in previous years, losing records of planting dates, or misjudging harvest readiness can affect their productivity and financial stability. To respond to challenges, farmers developed their own workarounds, such as taking photos with mobile phones and relying on timestamps, using informal verbal reminders with family members, and creating hand-drawn notes.

During our interviews, participants frequently mentioned that their farming layout was fragmented and complex because they cultivated a wide variety of crops across their farmland. This type of farming generates a large volume of detailed data and information, such as planting and harvest schedules, crop rotation history, soil and fertilizer records, pest management notes, and yield estimates for each plot. Recording and managing this information often takes a significant amount of time and can be inefficient and challenging without appropriate tools for tracking.

P7 said that the harvesting season is always filled with tasks, including weeding and checking every type of crop, and that they can barely find time for anything else, including keeping necessary records:

\begin{quote}
    \textit{So another thing—I guess you'd say a difficulty for me—is just time management. Because when you're planting, you also have to be de-weeding. And then once you're harvesting, you've got to check all kinds of your produce and everything. So you just have so much to do, you don't have the time on hand to help you get all those tasks done.} (P7)
\end{quote}

As the quote shows, during the season, there are always back-to-back tasks waiting for farmers, making it difficult for them to sit down and organize the necessary information. As a result, they often do not engage in formal information tracking practices. Another participant, P6, shared that while farming is physically demanding, the main barrier to engaging in any form of information tracking was the cognitive load—she simply did not have the mental capacity to keep up with it:

\begin{quote}
    \textit{Oh, because it's difficult—especially when I tell you that farming takes a lot of your time. That means we really don't have much time to remember, like—``Oh, okay, I need to record this, I need to record that...'' But we don't use any form, any app for it, because it's really hard for us to remember to think, ‘Oh, okay, I need to record this, I need to record that.} (P6)
\end{quote}

As P6 noted, she does not use existing information tracking tools (e.g., mobile apps) because the need to remember to record information adds another layer of cognitive load, making farming even more time-consuming. As a result, she is unable to engage in any formal tracking practices.

Moreover, participants mentioned that other factors added to the burden of their farming routines, beyond the challenge of growing many types of crops. One main factor was land rotation, where farmers had to divide their land and leave at least one part without planting anything so the soil could regain nutrients. P5 shared that he does land rotation regularly to keep the soil healthy, but said it makes planning the farm more difficult because it requires keeping track of which sections were planted, which were left out, and when each section needs to rotate. This adds to the complexity of record-keeping. This process often created challenges for participants in tracking the location of each crop and remembering their rotation plans, as they needed to refer to detailed farming records from previous years to plan for future seasons.

\begin{figure*}[h]
  \centering
  \begin{subfigure}[t]{0.48\textwidth}
    \centering
    \includegraphics[width=\linewidth]{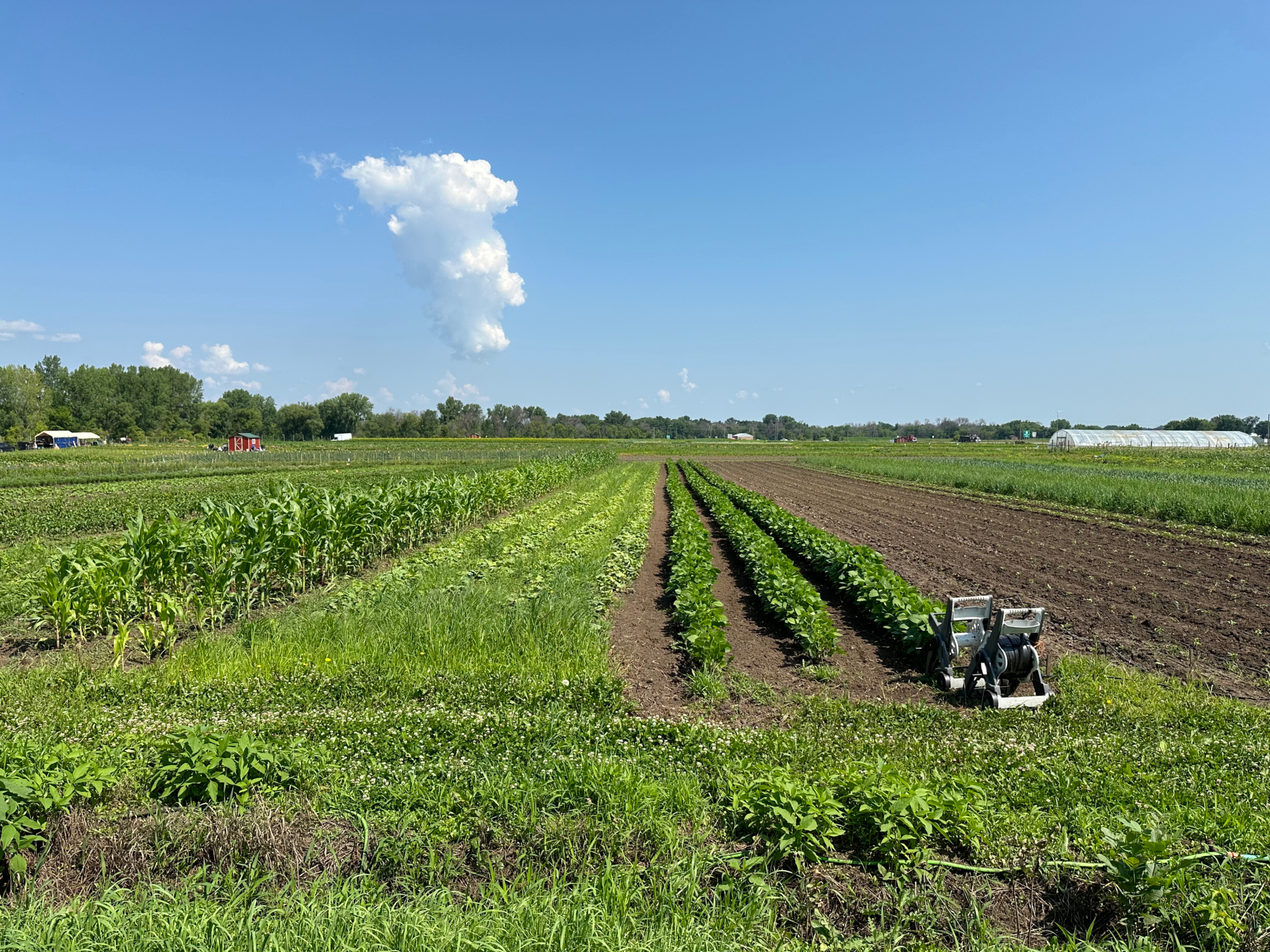}  
    \caption{The farm used intercropping practices, with crops on the right side just beginning to sprout, while those on the left side had already entered the growing stage.}
    \label{fig:inter}
    \Description{The farm used intercropping practices, with crops on the right side just beginning to sprout, while those on the left side had already entered the growing stage.}
  \end{subfigure}
  \hfill
  \begin{subfigure}[t]{0.48\textwidth}
    \centering
    \includegraphics[width=\linewidth]{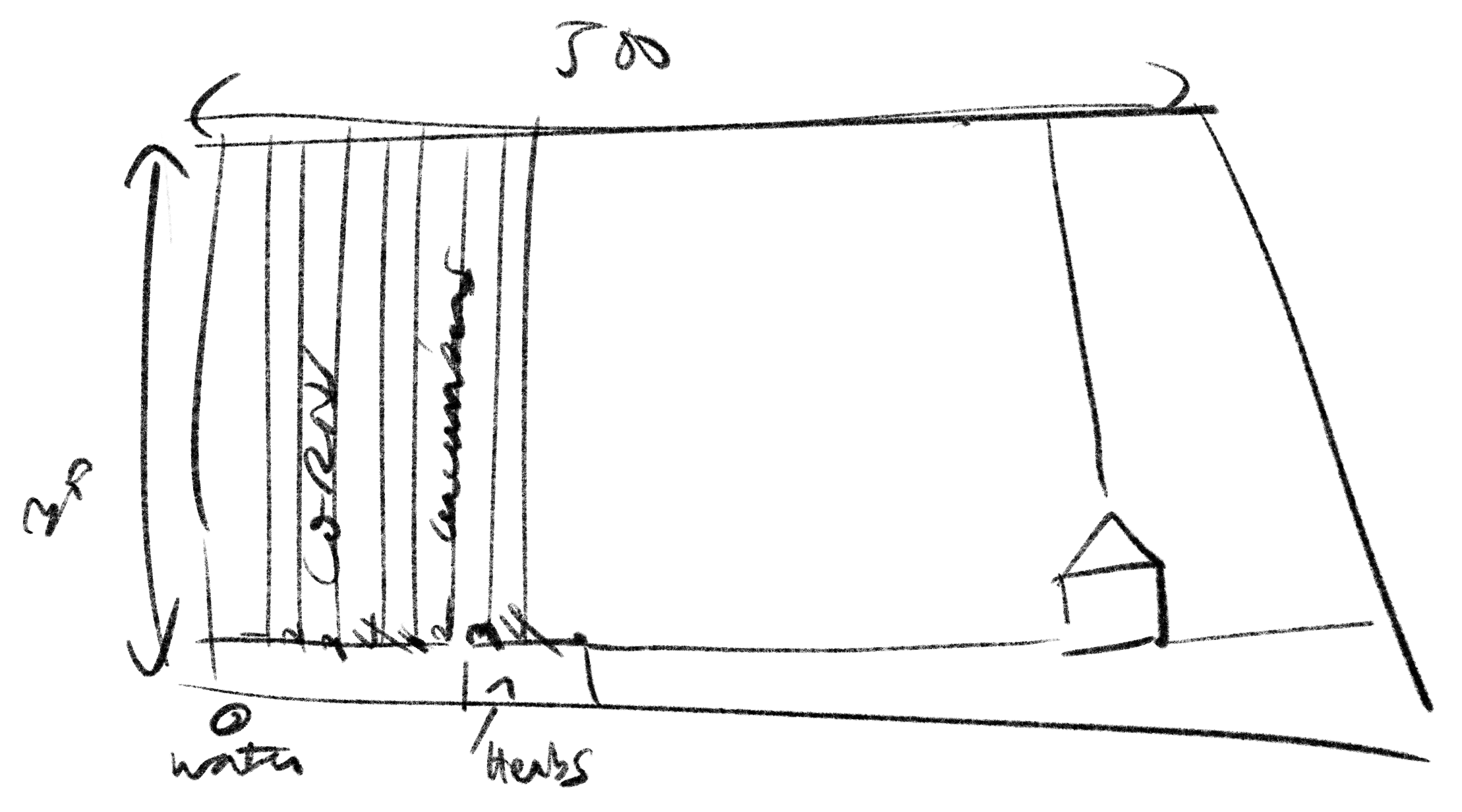}  
    \caption{During the interview, a farmer used the researcher's iPad to sketch out their farming plans for us.}
    \label{fig:sket}
    \Description{During the interview, a farmer used the researcher's iPad to sketch out their farming plans for us.}
  \end{subfigure}
  \caption{Intercropping and planning sketch.}
  \label{fig:combined2}
\end{figure*}

In addition to land rotation, participants share that they often cultivate crops with short growing cycles to increase yield. However, these short cycles require continuous and frequent farm planning, which makes information tracking more labor-intensive and complex (see Figure~\ref{fig:inter}). For example, P6 grows short-term crops, which require continuous planting and coordination with land rotation. She described the challenge of maintaining a steady supply while organizing planting schedules for the farmers' market.

%%\begin{figure}[h]
%%  \centering
%%  \includegraphics[width=0.6\linewidth]{Figure-intercropping.jpeg}  
%%  \caption{The intercropping practices on the farm. Crops on the right side of the farm were just sprouting, while those on the left side were already in the growing stage.}
%%  \label{fig:inter}
%%  \Description{The intercropping practices on the farm. Crops on the right side of the farm were just sprouting, while those on the left side were already in the growing stage.}
%%\end{figure}

To keep track of the information generated by their complex farming practices, many participants reported trying existing information tracking tools, such as Excel. However, they found these tools difficult to use due to high learning barriers. Participants shared that the tools were not straightforward enough to use independently, given their limited technology literacy. For example, P1 believed that Excel could be helpful for tracking, but he needed assistance from family or friends with stronger technology skills to complete the process:

\begin{quote}
    \textit{Maybe Excel could help us keep track of the farm… but my sister has to help me because she’s a lawyer. It’s not easy to use, but it would be very helpful. If you could create a form where we just fill in the numbers, then someone could help type it in. That would be easier.} (P1)
\end{quote}

The participant suggested having a tool with a simple input format that does not require extensive typing. In addition to Excel, another commonly mentioned tool was Google Maps, which some participants tried using to record crop locations. However, they found it “not precise enough.” Because participants grow a wide variety of crops on small plots of land (5 to 10 acres), with each crop occupying only two or three rows, the existing maps lacked the accuracy needed to distinguish crop types and their exact locations. As P5 shared, the tools do not provide the level of location detail he needed:

\begin{quote}
    \textit{I tried to use Google Maps and all that, but even with Google Maps, it’s not sophisticated enough. I can’t just map things out generally. When you try to compare what you see on Google Maps to what’s actually on the ground, it doesn’t work. You’d need tools that can tell you, ``From here to here, this is what’s on the map.'' You’d need some kind of navigation tool to tell you exactly where to start planting and how straight the rows should be, and so on. I’m not there yet.} (P5)
\end{quote}

The limitations of existing information tracking tools significantly hinder participants’ farming efficiency. As an alternative to tools that failed to meet the demands of their unique farming practices, participants developed their own workarounds. For example, P6 shared that she created a personal system to track planting times by taking photos with her mobile phone and using the timestamps to record dates:

\begin{quote}
    \textit{Sometimes I make sure to at least stamp a time and date, just to know, 'Okay, I planted these crops on this date.' That’s something I want to keep recording—when I planted something—so I can know how long that season lasted and when to start the next one. That’s one issue we have at the garden. We want to be able to know the timeline of when things grow so we can keep the crop cycle going all year.} (P6)
\end{quote}

P6 developed a workaround that would help her continuously track the timing of crop cultivation. Similarly, P5 created his own workaround to address the lack of tools with precise location data. He used hand-drawn maps of his farmland to record key details, including the number of rows, the type of crop on each row, and the locations of irrigation equipment and tractors (see Figure~\ref{fig:sket}):

\begin{quote}
    \textit{When I do it, I just draw on paper—corn here, beans or squash there. It’s not exact, just an overview. Since each row is four or five feet apart, I try not to plant the same crop in the same place the next year. I’ll note where the herbs are, where the water is, the little house, the tractor rows, and estimate distances—like 300 feet here, 500 feet there. Then I mark rows for crops, like four rows of corn, and maybe four more for cucumbers.} (P5)
\end{quote}

%%\begin{figure}[h]
%%  \centering
%%  \includegraphics[width=0.6\linewidth]{FIgure-Planning Sketch.png}  
%%  \caption{A farmer drew for us their sketch of farming plans using an iPad during the interview.}
%%  \label{fig:sket}
%%  \Description{A farmer drew for us their sketch of farming plans using an iPad during the interview.}
%%\end{figure}

This example shows how P5 used hand-drawn maps as a practical workaround to visualize the farm layout, manage crop placement, and support rotational planning when digital tools lacked the necessary precision.

In summary, our findings highlight key characteristics of farmlands managed by Hmong farmers, including diverse crop types, complex layouts, and continuous farming routines. These features made existing information tracking tools (e.g., Excel, Google Maps) ineffective for their needs. Instead, the farmers developed their own workarounds, such as hand-drawn maps and timestamped photos, which were more practical and better aligned with their farming practices.

\subsection{Relying on Hand Tools in the Absence of Equipment for Diverse Small-Scale Farming}

\begin{figure*}[h]
  \centering
  \begin{subfigure}[t]{0.48\textwidth}
    \centering
    \includegraphics[width=\linewidth]{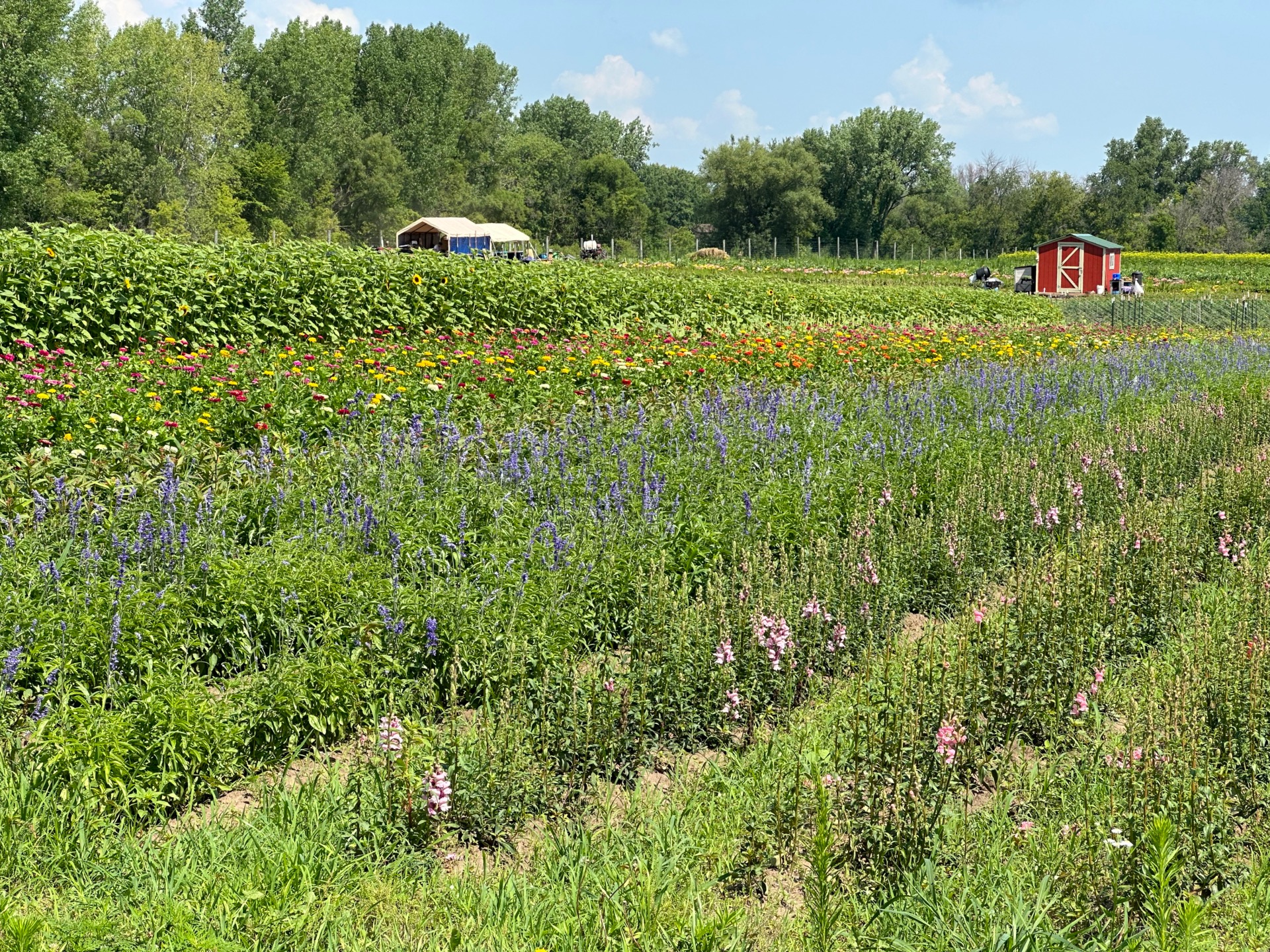}  
    \caption{Crops were planted in dense layouts.}
    \label{fig:dense}
    \Description{Farmers' crops were planted in dense layouts.}
  \end{subfigure}
  \hfill
  \begin{subfigure}[t]{0.48\textwidth}
    \centering
    \includegraphics[width=\linewidth]{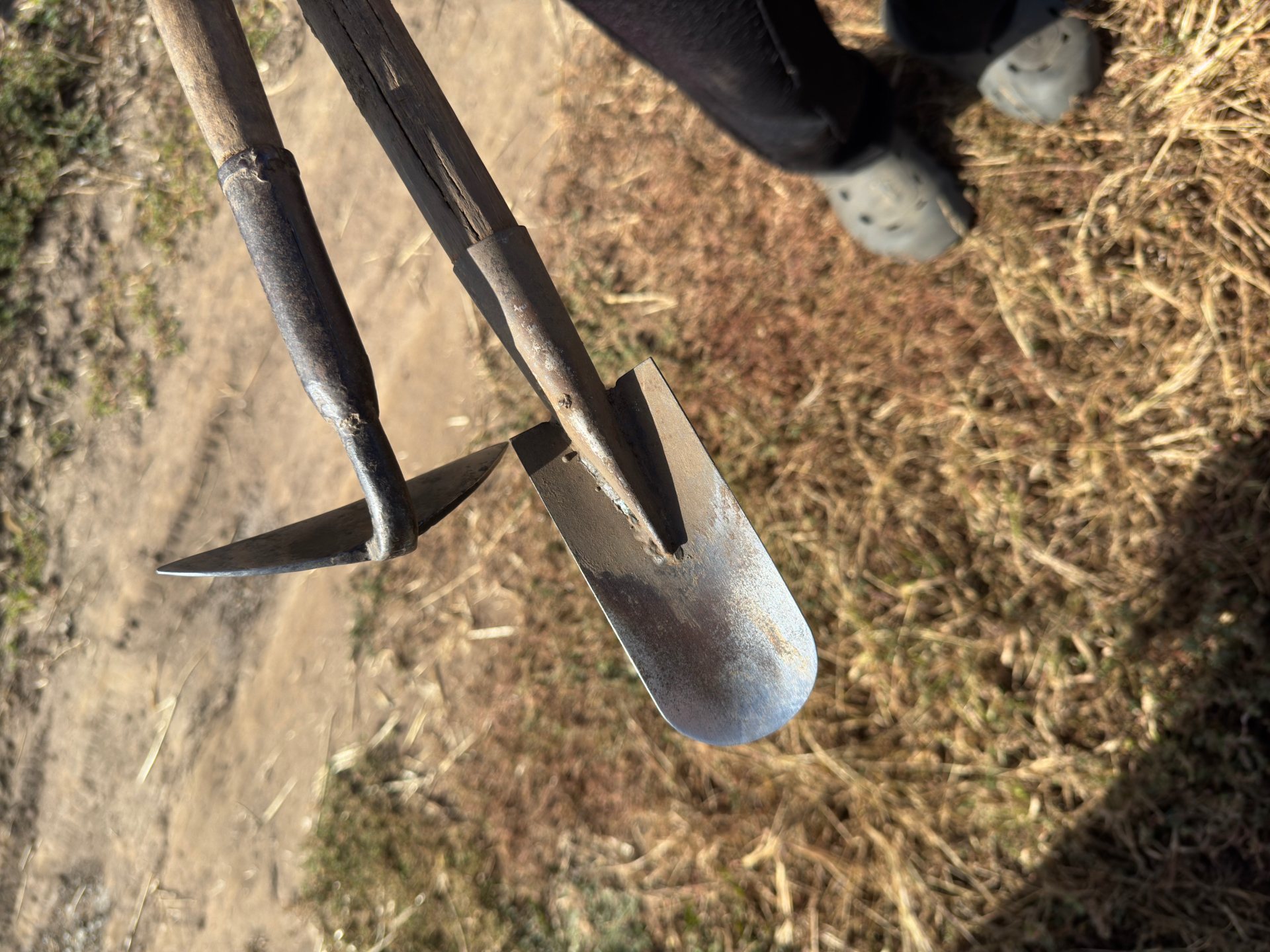}  
    \caption{A farmer showed us their mostly used hand tools, including a traditional hoe and a shovel.}
    \label{fig:hand}
    \Description{A farmer showed us their mostly used hand tools, including a traditional hoe and a shovel.}
  \end{subfigure}
  \caption{Dense layout and hand tools.}
  \label{fig:combined3}
\end{figure*}

According to participants, one key challenge in their farming practices is the mismatch between their needs and the design of standard farming equipment. Hmong farmers lease small plots (5 to 10 acres) from nonprofit organizations and cultivate a wide variety of crops within these limited areas. This results in exceptionally dense layouts that are difficult to navigate with conventional machinery (see Figure~\ref{fig:dense}). In addition, participants frequently mentioned unpredictable weather, which further complicates their ability to use existing equipment effectively.

%%\begin{figure}[h]
%%  \centering
%%  \includegraphics[width=0.6\linewidth]{Figure-Dense Plot.jpeg}  
%%  \caption{Farmers' crops were planted in dense layouts.}
%%  \label{fig:dense}
%%  \Description{Farmers' crops were planted in dense layouts.}
%%\end{figure}

To work in their densely planted fields, participants emphasized the need for equipment that is smaller and more accurate. For example, tractors must be able to move carefully between rows of different crops without causing damage. However, most existing equipment is designed for large-scale farms that grow only a few types of crops. Farmers operating large farms have more space between rows and simpler layouts, which makes it easier for tractors to navigate. As P10 noted, larger farms offer uniform layouts that simplify machine use:

\begin{quote}
    \textit{Unlike the tractors those white farmers have... they just go straight, looking forward, because everything is the same. They know where they're going.} (P10)
\end{quote}

In contrast, Hmong farmers often rely on hand tools, which, while effective in tight spaces, are labor-intensive and inefficient (see Figure~\ref{fig:hand}). Similarly, P6 shared her dissatisfaction with the lack of precision in tractor use. She explained that tractors often failed to complete tasks as expected, such as spreading fertilizer or digging holes. As a result, she frequently had to redo the work:

\begin{quote}
    \textit{Sometimes it's not very accurate. You can't fully put the dirt on there, so they always have to go back and fix it. Or when fertilizer is poured into the tractor, it might pour too much or not spread evenly. And using the tractor itself limits you when it comes to the small details. You can use it for the big stuff, but there's a limit to what it can do.} (P6)
\end{quote}

This quote illustrates that Hmong farmers are highly detail-oriented in their farming tasks and expect every small detail to be handled with care, which tractors often fail to achieve.

%%\begin{figure}[h]
%%  \centering
%%  \includegraphics[width=0.6\linewidth]{Figure-Farmer hand tool.jpeg}  
%%  \caption{A farmer showed us their mostly used hand tools, including a traditional hoe and a shovel.}
%%  \label{fig:hand}
%%  \Description{A farmer showed us their mostly used hand tools, including a traditional hoe and a shovel.}
%%\end{figure}

It is not only the small scale of farmlands that prevents existing farming equipment from effectively supporting Hmong farmers’ routines, but also the limited adaptability of this equipment to varying weather conditions. Participants frequently described rainy weather as a major obstacle that disrupts their work. Due to the complex and densely planted layouts of their farms, they often cannot pause farming activities even during bad weather. Farmers emphasized that most equipment, especially tractors, is not well-suited for use in wet conditions. P3 shared that tractors become even less accurate in the rain and can damage both crops and soil when entering the fields during wet weather:

\begin{quote}
    \textit{The only disadvantage of using the machine is when it rains. When it rains, it runs over everything, so you can't go in with the machine. If you have your vegetables there and it rains, the tractor will pull the plants out and create deep ruts in the soil.} (P3)
\end{quote}

As this quote shows, participants viewed the adaptability of farming equipment to different weather conditions as one of the most crucial factors. They noted that rainy weather significantly increases their workload due to the lack of tools suited for wet or covered fields. Farmers mentioned that methods like mulching and building low tunnels are necessary to protect crops from excess water, but these make follow-up tasks like weeding more difficult. Since tractors cannot operate on wet or obstructed ground, farmers must rely on hand tools as the only viable option for completing many tasks. However, this is especially difficult due to limited labor availability.

According to P1, tilling one acre of land with hand tools, such as a tiller, can take an entire day, while a tractor could complete the same task in just 30 minutes:

\begin{quote}
    \textit{Hand tools are only suitable for less than two acres, maybe one acre. Beyond that, you need a tractor. One acre is okay, but a tiller is too small—it'll take forever. You could do one acre in 30 minutes with a tractor, so it saves a lot of time. That's the problem for Hmong farmers. It's not that they want to work every day, but when you're only using your hands, it takes all day just to do one acre.} (P1)
\end{quote}

As this quote illustrates, hand tools may be sufficient for very small plots, but for farms ranging from 5 to 10 acres, their efficiency is extremely limited.

To address the challenges posed by ineffective farming equipment, participants shared their expectations and ideas for the tools they hoped to own in the future. For example, P11 described several types of equipment she believed could significantly improve her current farming routines, including planters and weed controllers. She believed that having the right equipment would improve efficiency, reduce labor, and, more importantly, allow her to continue farming within the community for a longer period of time.

This subsection highlights that existing farming equipment lacks the adaptability needed for the small-scale and diverse farms managed by Hmong farmers. Hand tools remain the only viable option, although they are labor-intensive and inefficient. 

\subsection{Community as a Multi-faceted Resource: Strengthening Economic, Emotional, and Knowledge Networks}

Our results showed that many farmers are affiliated with local farming associations that help bridge the gap between the community and external systems. These associations function as collectives, providing access to essential resources such as land, capital, and farming equipment. Since many farmers primarily speak Hmong, limited English proficiency can make it difficult to access updated farming information or participate confidently on English-dominant platforms. Nonetheless, participants relied on the strength of their social networks, forming close-knit communities that supported knowledge exchange, resource sharing, and adaptation to their new environment.

A key barrier frequently identified by participants was limited access to farmland. Purchasing and owning farmland was described as extremely difficult, often requiring years of effort. Participants attributed these challenges to high land prices, limited access to credit, and a lack of existing land or financial assets. For example, P1 shared his ongoing efforts to purchase land, noting that many Hmong farmers are unable to afford farmland in the U.S.:

\begin{quote}
    \textit{The Hmong people are poor too… I plan to buy my own farm. Yeah. But it's going to be a couple of years… It’s just that, the thing is, Hmong farmers are the same no matter where we go — they don't have land… That's their biggest problem. Unless they're already rich and they own a lot of land... But you’re an average farmer.} (P1)
\end{quote}

He mentioned that he owns more farming equipment than most Hmong farmers, which he explained was a result of obtaining organic certification early in his career. This achievement helped increase his income. Even so, he had only recently begun planning to purchase land.

Participants also emphasized that limited English proficiency created challenges beyond land access, particularly in acquiring or sharing up-to-date farming information on English-dominant platforms. While issues like land access and equipment reflect material barriers, participants also experienced less visible challenges, such as language-related barriers that limited communication and access to information.

To navigate these challenges, participants highlighted the importance of joining local Hmong farming communities as a way to build support networks. These communities provided access to farmland and equipment rentals, and supported members in applying for farming-related grants, which often required knowledge and language proficiency.

\begin{figure}[h]
  \centering
  \includegraphics[width=0.48\linewidth]{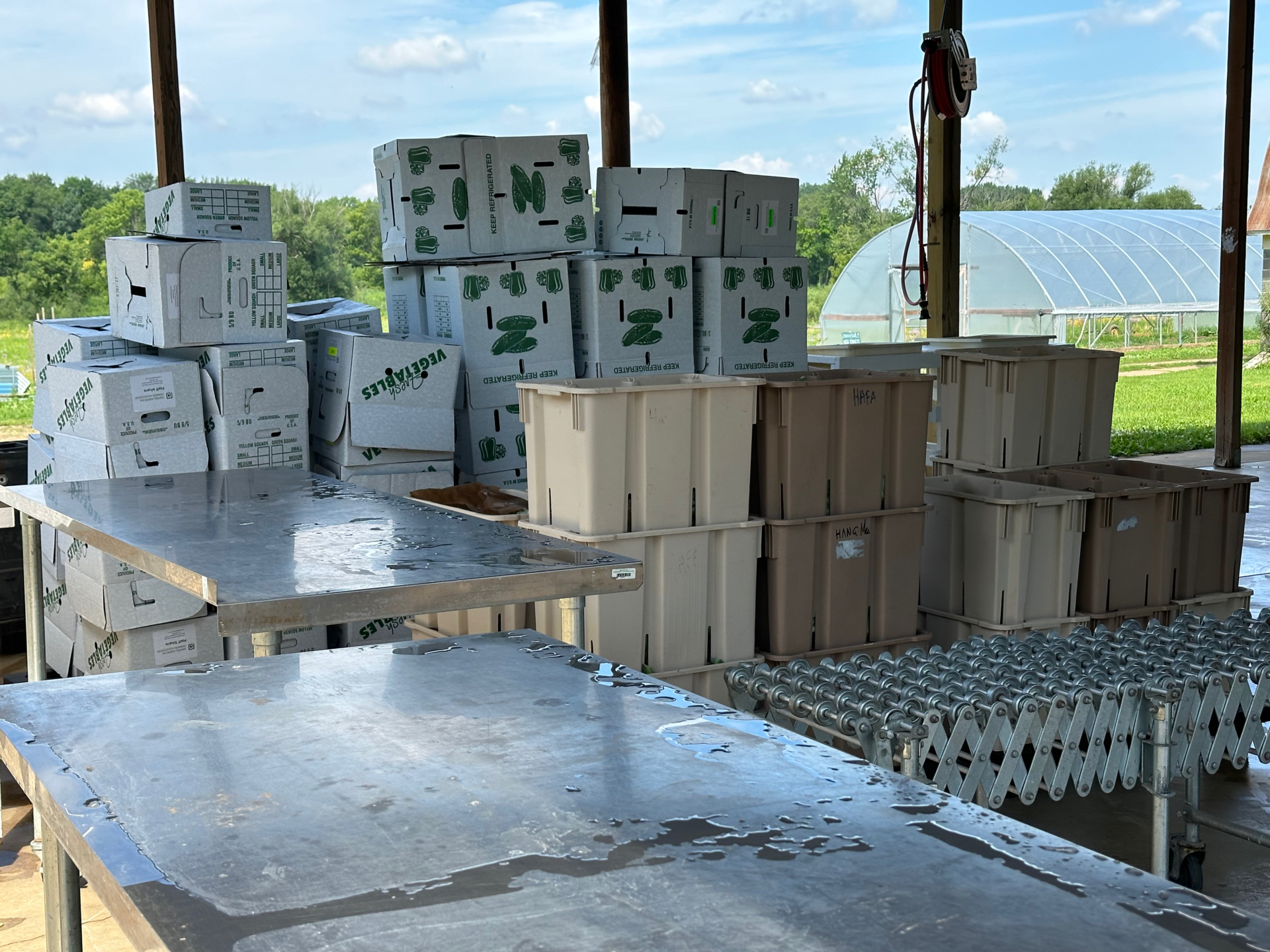}  
  \caption{The staff members of a local association packaged farmers’ produce outside the cold storage on the farm, which would be delivered to the customers enrolled in the Community Supported Agriculture (CSA) program.}
  \label{fig:pack}
  \Description{The staff members of a local association packaged farmers’ produce outside the cold storage on the farm, which would be delivered to the customers enrolled in the CSA program.}
\end{figure}

In addition to direct support from the community (e.g., farmland, equipment, grants), participants also noted that their community helped strengthen local economic networks. These networks provide opportunities of connections with other communities and farmers’ markets (see Figure~\ref{fig:pack}). P10 shared that the community helped him expand his market by connecting with various organizations:

\begin{quote}
    \textit{And with that, we can have a better source of marketing—maybe wholesale, Community Supported Agriculture, or through the school. That’s something the organization helps students with, so we can generate income from it instead of having to go through all the steps ourselves. The organization acts as a third party, connecting us to various sales channels beyond just the farmers’ market.} (P10)
\end{quote}

He also mentioned that the community’s positive reputation benefited his efforts to build new connections. When introducing himself during marketing activities, people immediately recognized the name of the community he belongs to. This familiarity increased customer trust at the farmers’ market, as many were already aware of the community’s good standing.

According to the participants, the community offers emotional support, which often motivates members to take concrete actions to help one another, such as donating food or money and checking in on each other's well-being. During the interviews, participants shared that they developed a strong sense of responsibility and showed a caring attitude toward other community members as they work as farmers associated with their local network. Several participants described making regular donations to support those in need. For example, P11 shared that she donates money and food to local communities on a regular basis:

\begin{quote}
    \textit{So every Saturday or Sunday, if I’m not at the market, I give my produce to the donation center so they have good food or produce to share with people in need. Also, I support the community because when I was still working, I would donate at least \$50 or so to an organization that helps provide for those in need.} (P11)
\end{quote}

The definition of community often extends beyond its immediate farming network. Participants noted that supporting the health and well-being of local residents is one of their core responsibilities. For example, P4 believes that providing fresh, healthy food with easy access is her primary responsibility as a community farmer:

\begin{quote}
    \textit{Fresh food—being able to provide the community with fresh food, being able to help the community stay healthy or support their health. I forgot what it’s called, but we sell some of the food to local companies that provide it to people who need it, they don’t go hungry. Basically, people who live in poverty and rely on some help with fresh, local vegetables because maybe they can’t afford that.} (P4)
\end{quote}

These examples show that the community plays an important role in building relationships among member farmers and local residents. Through trust, care, and shared practices, farmers naturally develop a sense of mission and responsibility—not only toward their close-knit community but also toward the broader society they are part of.

Similarly, participants emphasized that the community enables the transmission of farming knowledge, particularly tacit knowledge that is difficult to express verbally, such as hands-on skills and experience gained through observation. Many shared that they first learned farming by following and observing senior community members during their early years. The community still serves as an essential source for exchanging farming strategies and perspectives. Because English-dominant information and resources are often less accessible, the community offers a more viable and trusted space for knowledge sharing for each other. For example, P7 described how community members openly discuss successful practices:

\begin{quote}
    \textit{So it's more like community knowledge, in a way... Or if we know someone who does that and has a good yield, we could ask, ``What did you do this year? Why is your stuff looking better than mine?'' Then they'll pass on the knowledge.} (P7)
\end{quote}
	
As this quote shows, Hmong farmers view the community as a crucial source for exchanging effective farming knowledge and perspectives with one another.

This subsection shows that forming a community is one of the most important ways Hmong farmers support their efforts to thrive in the U.S. They view their community as a vital source of resources that helps members overcome challenges, such as difficulties in accessing farming information from English-dominant source and barriers to obtaining essential resources like land, capital, and equipment. The community provides opportunities for financial assistance, emotional support, and knowledge sharing.

\subsection{Envisioning Intergenerational Collaboration as a Way to Make Farming Careers More Expandable and Sustainable}

The study results show that, even though farming was identified as an essential part of life that participants wish to preserve, there are significant generational differences in perspectives toward it. During the interviews, participants emphasized the need for farming to evolve with changing times. They discussed challenges in presenting the positive aspects of farming in effective ways, including the use of appropriate knowledge and tools. They also highlighted the potential benefits of leveraging the younger generation's technological literacy and English proficiency.

Participants described a typical Hmong farming family as consisting of three generations. The older generation, in their 60s to 80s, sees farming as essential for sustaining life and honoring cultural values. The middle or "sandwich" generation, in their 30s to 50s and children of the older generation, often feel a strong sense of responsibility to continue farming. Those in their early 30s, who grew up helping in the fields, shared that they continue farming despite low financial returns because they feel a duty to carry on the family legacy. In contrast, the younger generation, under 30, shows little interest in farming due to limited opportunities and low motivation. They often view it as physically demanding and economically unrewarding. As a result, the older generation struggles to engage them. This generational divide reflects a shift in how farming is perceived.

For example, P5 shared that his children refused to learn farming skills from him, describing the work as backbreaking and expressing frustration with the harsh weather conditions: 

\begin{quote}
    \textit{My kids have helped me on and off since they were little, but they hate it. They were never really taught, and I didn’t teach them properly because I didn’t know how either. It’s backbreaking work, and the weather—hot, humid, then cold—makes being outside for over eight hours really hard on their mental health. They just don’t like it.} (P5)
\end{quote}

Similarly, P6 noted that the younger generation views farming as labor-intensive and tiring, showing limited interest in pursuing it as a career. However, farmers strongly emphasized the potential benefits of involving the younger generation in farming. They noted that younger people bring strengths that older generations may not possess, such as the ability to use social media and speak fluently in English with a wide range of customers. These skills are becoming increasingly important for helping farming businesses grow in step with broader developments. Farmers also observed that younger generations are more receptive to new ideas and methods.
 
Participants noted that involving young family members in the sales procedures helps cultivate their communication with customers, as well as enhancing their knowledge about crops. Engaging them in these critical tasks also potentially fosters their interest in engaging in farming collaboratively.
 
\begin{quote}
    \textit{Our younger siblings, nephews, and nieces always come to help us. The younger generation may not like hard labor, but they help at the farmers' market, which builds useful skills. They learn to communicate with the community and recognize the vegetables. When a customer asks, “What is this vegetable?” they can explain it and suggest ways to cook it. That knowledge becomes ingrained—it doesn’t feel like “work” to them.} (P6)
\end{quote}
 
During our interviews, younger participants shared childhood stories of working alongside their parents and grandparents, describing how farming gradually became a meaningful family memory and part of their cultural experience. For the younger generation, these early experiences provided both background knowledge and a sense of farming as a hobby, which later became an important part of their personal identity. This connection often motivated them to return to farming after exploring other career paths. As P4 noted, farming feels deeply rooted in their blood and bones, shaped by collective family activities in the fields.
 
\begin{quote}
    \textit{The oldest daughter worked on the farm with my grandparents. They’d go every morning and come back before dark. My mom stayed home to raise the kids and didn’t get much farming experience until she was older and the younger ones could take care of themselves. That’s how I see myself—I’m Hmong, and I come from a farming background. In a way, it feels like it’s in my genes, in my blood.} (P4)
\end{quote}
 
By naturally engaging in family-based activities, farming has been a part of P4's life rather than something that needed to be formally taught or consciously chosen.
 
\begin{quote}
    \textit{It’s more of a lifestyle. You learn it from your parents, and as you grow older, you just know how to do it. Then, with the next generation, you hand them the tools and say, “Do the same thing.” You learn by doing, and the knowledge grows with you.} (P4)
\end{quote}

\begin{figure}[h]
  \centering
  \includegraphics[width=0.48\linewidth]{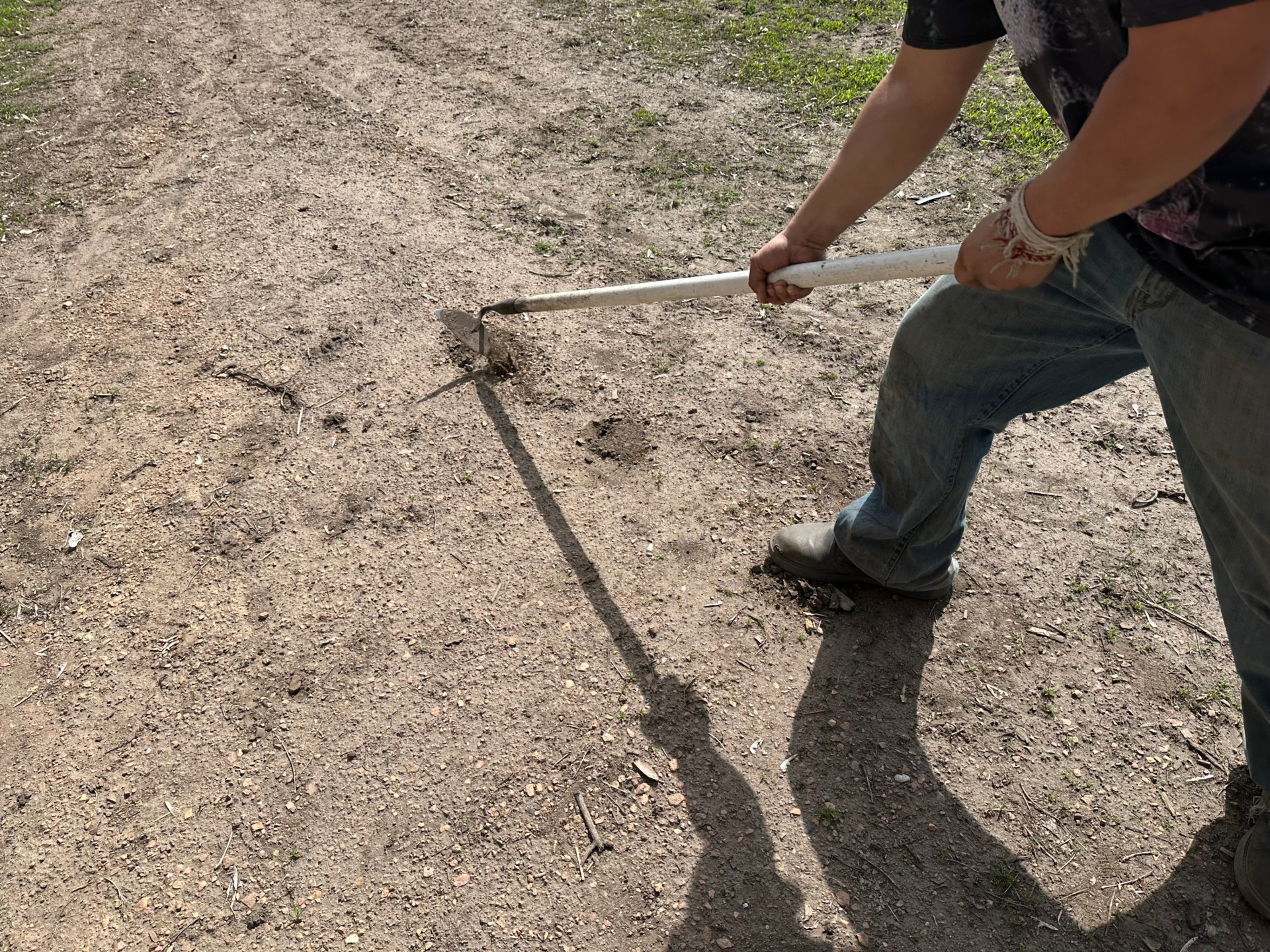}  
  \caption{A farmer showed the study team how their parents taught them to measure the spacing between planting holes by gripping the hoe handle at a specific spot, using their hand placement as a reference.}
  \label{fig:teach}
  \Description{A farmer showed us how their parents taught them to measure the spacing between planting holes by gripping the hoe handle at a specific spot, using their hand placement as a reference.}
\end{figure}
 
During our interviews, participants acknowledged that their children were not effectively taught farming skills due to a lack of formal training. Many farmers had learned these skills informally through everyday participation with their own parents, which often resulted in inconsistent and less effective practices (see Figure~\ref{fig:teach}). Participants expressed the need to integrate more formal farming methods with precise controls when training the younger generation.
 
\begin{quote}
    \textit{I learned a lot from my mother about how to farm. But the problem was, our methods weren’t sustainable, and they didn’t plant properly. They used to scatter seeds everywhere, which made weeding really difficult. But if you plant in straight rows, it’s easier to till the soil and manage weeds. Since coming here, we’ve learned how to improve—like saving seeds for the next year and choosing the best ones.} (P1)
\end{quote}

As noted by P1, traditional training methods have their limitations and can be enhanced by incorporating more legitimate training, which shows potential for improving learning experiences for younger generations. 

In this subsection, we explore the differing views on farming between generations within Hmong farming families and discuss their suggestions for engaging the younger generation in farming activities. Although the older generation hopes that the younger generation will continue the farming business, younger people often find farming unappealing, viewing it as labor-intensive with low returns. To improve this perception, farmers recommended different ways to enhance training and motivation for the next generation.

\section{Discussion}
\subsection{The Critical Role of Bonding Capital in Community-Based Small Farms}

In our study, Hmong farmers developed their own strategies for sustainable and effective farming, such as timestamped photos and hand-drawn maps. When intentionally connected and shared within the community, these practices have the potential to foster community capital. As Yosso~\cite{yosso_whose_2005} highlighted, such practices exemplify forms of navigational, familial, linguistic, and social capital. Similarly, Mathie and Cunningham~\cite{mathie_clients_2003} emphasized that these practices reflect forms of individual capital that communities can map, link, and mobilize.

In many resource-scarce contexts, individual capabilities can be cultivated within community networks and, in turn, become resources that sustain collective well-being. Prior literature in critical race theory and community development~\cite{yosso_whose_2005,mathie_clients_2003} has shown that individual skills (e.g., language) can form the foundation of community capital when strengthened through social affiliations and personal networks. Our study revealed that farmers were already informally mobilizing their individual capital. For instance, they shared workarounds and practical knowledge within their families. When aggregated and mobilized, these individual forms of capital became evident as three forms of social capital within the Hmong farming community.

Such practices align with the well-established distinctions of bonding, bridging, and linking capital~\cite{woolcock_SocialCapitalEconomic_1998,putnam_BowlingAloneCollapse_2001}. Sharing successful farming experiences with other Hmong farmers reflects bonding capital (the strong ties that connect similar individuals within a group). Establishing new sales channels by connecting with farmers markets and schools illustrates bridging capital, which relies on weaker ties across diverse groups. Finally, attending farming training programs offered by local agencies represents linking capital, since it connects community members with external institutions and formal knowledge systems. This aligns with previous studies on social capital in HCI and CSCW~\cite{kokkalis_etal_FounderCenterEnabling_2017, metaxa-kakavouli_etal_HowSocialTies_2018, leal_etal_GrowingTogetherRemaining_2021}. For example, Lampe et al. examined how using Facebook to expand personal networks can enhance social capital, and found that frequent, meaningful engagement with the platform was particularly effective in strengthening bridging capital, such as connections to broader networks and diverse social groups~\cite{lampe_etal_UsersNonusersInteractions_2013}. Bonding capital, or strong and close ties within existing networks, was also enhanced, though to a lesser extent. Similarly, Gray et al. found that individuals’ ability to expand knowledge and access new information through social media was positively associated with bridging capital~\cite{gray_etal_WhoWantsKnow_2013}. Although bridging capital has been actively studied in HCI and CSCW, particularly through technologies that facilitate the expansion of social networks~\cite{lampe_etal_UsersNonusersInteractions_2013, hsiao_dillahunt_PeoplenearbyApplicationsHow_2017, burke_kraut_UsingFacebookLosing_2013}, the role of bonding capital remains less well understood. A few exceptions exist in studies that explore family-centered technologies designed for specific, contextualized goals such as healthcare~\cite{pina_etal_DreamCatcherExploringHow_2020}. This gap is particularly relevant to our context, where bonding capital emerged as the most prominent and accessible form of social capital among community members. Furthermore, while bonding, bridging, and linking capital have each been examined in HCI and CSCW, their interrelationships remain underexplored. Investigating how these forms of social capital can support and reinforce one another is critical. For instance, bonding capital—grounded in trust and close community ties—may serve as a foundation for developing both bridging and linking capital over time.

Our findings support this perspective. Among the three types of social capital, Hmong farmers most heavily relied on bonding capital (i.e., community networks rooted in close-knit family ties and mutual support) to navigate challenges. They viewed these relationships as essential to sustaining their farming practices, including sharing updated information, lending tools, and passing agricultural knowledge to younger generations. Importantly, we found that bridging and linking capital often emerged from these strong internal networks. For example, farmers became involved with nonprofit organizations through personal connections, which in turn helped them access broader networks such as farmers’ markets, customer groups, and schools (bridging capital), as well as institutional resources like universities and government agencies (linking capital). Building on prior studies that emphasize the importance of social capital in navigating challenges, our findings reveal how bonding capital can serve as a foundation for expanding social ties. Strengthening these connections is especially important for promoting more sustainable agricultural practices.

We suggest that HCI researchers consider the distinct forms of social capital present in each community and explore ways to strengthen them to address local challenges. In our study of Hmong farmers, bonding capital played a foundational role in supporting both bridging and linking capital, though this dynamic may differ in other contexts. Bonding capital can inform the design of technologies that foster trust and peer support within close-knit groups. Bridging capital can inspire tools that connect community members across different networks to facilitate the exchange of information, resources, and practices. Linking capital can guide the development of platforms that support communication and coordination with institutions, nonprofits, or customers through various market channels. Our findings also show how farmers’ individual skills and workarounds, once shared, became valuable resources for the broader community. Future research in HCI can build on this by designing technologies that help formalize such exchanges, enabling individual capabilities to contribute more directly to community-level capital.Taken together, these dimensions offer a valuable framework for designing contextually grounded and sustainable social technologies for small communities. In the following section, we outline specific design opportunities informed by this insight.

\subsection{Designing Technologies for Community-Based Small Farms}
\subsubsection{Intergenerational Collaboration to Find Opportunities for Sustainable Practices} \hfill

\noindent From our interviews, we learned that farming is more than a livelihood for Hmong farmers; it is closely tied to their lifestyle, heritage, and cultural identity. Community members use farming as a way to pursue educational and cultural goals, such as passing traditions to younger generations and maintaining farming practices. Community-based small farms face persistent challenges, including weather variability (freezing, drought, excessive rain)~\cite{bakhsh2019adaptation} and sustainability demands such as land rotation and the preservation of farming heritage. In addition, technological support for these communities remains limited~\cite{mhlanga2023digital}. 

Despite these constraints, we found that such farms have developed resilient practices to adapt. Unlike large-scale agricultural operations, small-scale urban farms—especially those that are community-based or immigrant-led—rely heavily on informal relationships and collective knowledge to sustain their everyday work (bonding capital)~\cite{cofre2019combinations}. Technology, therefore, should aim to strengthen these existing social systems and support ongoing practices, rather than replace them. One key challenge that emerged, however, is the difficulty of passing accumulated farming experience and cultural heritage to the next generation. While farmers acknowledge the value of intergenerational collaboration, many struggle to implement it effectively in practice.

HCI and CSCW literature has explored intergenerational collaboration in various domains such as physical activity, education, and healthcare~\cite{hunt_etal_ChildrenUsingTabletop_2025, yuan_etal_UnilateralDominanceCollaborative_2024, binda_etal_PhamilyHealthPhotoSharing_2018, liaqat_munteanu_SocialLearningFrameworks_2019, shin_towards_2019}. These studies have identified both barriers and opportunities for technology-mediated intervention. For example, Hunt et al. suggested that having a coordinator to provide support and mediate conflicts can enhance intergenerational communication and collaboration~\cite{hunt_etal_ChildrenUsingTabletop_2025}. Similarly, Pina et al. identified privacy concerns in the context of intergenerational health data sharing~\cite{pina_etal_PersonalInformaticsFamily_2017a}. These concerns included children’s fear of losing autonomy over their schedules (e.g., revealing late-night habits) and parents’ fear of children misusing the data (e.g., unnecessary competition among family members). The results indicated that giving each generation greater control over their own data can help mitigate these concerns. Shin et al. also noted that although different generations tend to have varying expectations regarding the frequency and mode of communication, empowering children to initiate conversations and express themselves may offer new opportunities to improve intergenerational communication~\cite{shin_etal_DesigningTechnologiesSupport_2021}.

Building on previous studies that have identified factors influencing intergenerational collaboration~\cite{cha_etal_CollaborativeHealthtrackingTechnologies_2025, chowdhury_etal_ListeningTogetherApart_2021,butzer_etal_GrandtotemSupportingInternational_2020}, we suggest that HCI researchers adopt a contextually grounded approach that considers each generation’s unique cultural background and strengths. In small-scale urban farming communities, this includes strong bonding capital among community members, close-knit family relationships, and intergenerational collaboration rooted in farming heritage. These social factors influence how technologies are adopted, shared, and sustained, and should be carefully considered when designing tools to support intergenerational collaboration in these contexts.

We propose designing tools that build on existing social capital while transforming individual innovations into shared community assets. In our study, farmers created information-tracking workarounds (e.g., timestamped photos, hand-drawn maps) and adapted hand tools in resourceful ways. Low-cost, adaptable technologies that support intergenerational collaboration could formalize and extend such practices. For example, a group platform might enable younger members to access elders’ farming knowledge without heavy labor, while contributing their own strengths such as translation or product promotion. In this way, individual workarounds and practical skills become resources that sustain and strengthen collective social capital.

While these examples highlight the mobilization of knowledge and everyday practices, another main challenge for Hmong farmers lies in gaining access to appropriate farming tools. Establishing platforms for crowdsourcing information on advanced tools could present a promising opportunity for younger generations, given their familiarity with technologies such as the internet and social media. Previous studies have identified a range of robotic farming tools already available on the market, including weed controllers and detectors, targeted spraying machine, data collection tools, and harvesting robots~\cite{aravind_etal_TaskbasedAgriculturalMobile_2017, shamshiri_etal_ResearchDevelopmentAgricultural_2018}. For example, farmers could access a shared platform that lists advanced tools that are not easily purchasable at the individual level and reserve them for a certain period. The platform could be operated by younger generations while encouraging participation from older farmers, and it could provide simple, verbally delivered instructions to help users operate each tool. When a group of farmers needs a new tool or specific knowledge, they could collaboratively access these resources and decide whether to borrow, share, or collectively invest. The platform itself could use simple mobile or web interfaces with voice guidance, multilingual prompts, and basic inventory tracking to support farmers with varying levels of digital literacy. By leveraging strong bonding capital to share information and by developing community-driven models such as tool sharing, rentals, or group purchasing, small-scale farming communities could improve access to advanced technologies and begin to address the persistent challenge of equipment shortages.

\subsubsection{Designing Low-barrier Information Tracking Tools for Collective Knowledge in Community-based Farms} \hfill

\noindent In our study, farmers shared strategies they use to adapt their planning and information-tracking practices to small-scale farming in their new country. As there is a lack of information management tools tailored to the needs and constraints of small-scale farming groups, farmers have developed workaround approaches to information tracking to manage these challenges. These approaches include taking photos with mobile phones and relying on timestamps, using verbal reminders with family members, and creating hand-drawn notes. Such practices inform the design of technologies that support farmers facing similar challenges of low digital-literacy.

Our participants’ practices are aligned with previous studies in HCI and CSCW that have explored information tracking and record-keeping in low-literacy populations, including children~\cite{jones_etal_KidKeeperDesignCapturing_2017, heshmat_etal_FamilyStoriesAsynchronousAudio_2020}, immigrants~\cite{wong-villacres_etal_ParentingActornetworkLatino_2019}, and older adults~\cite{harrington_etal_EngagingLowIncomeAfrican_2019}. For these populations, limited digital-literacy can lead to low efficiency and impede their access to advanced digital tools and valuable information, such as limited access to healthcare resources and worse healthcare outcomes~\cite{harrington_etal_EngagingLowIncomeAfrican_2019}. To mitigate such barriers, these studies have proposed simple and accessible input methods such as haptic feedback~\cite{cingel_piper_HowParentsEngage_2017}, voice input~\cite{danovitch_etal_ChildrensUnderstandingUse_2025, beneteau_etal_CommunicationBreakdownsFamilies_2019}, drawing interfaces~\cite{zhang_etal_ObserveItDraw_2023}, using storytelling and metaphor~\cite{hunt_etal_MakingMetaphorSandwich_2024}, and touch-based interactions~\cite{yarosh_abowd_MediatedParentchildContact_2011}. 

These studies showed how technologies could be designed based on participants' literacy levels, keeping functions simple and intuitive, and adapting alternative interaction paradigms to users’ mental capabilities. For instance, Harrington et al. suggested using simple numbers and intuitive color codes when designing fitness apps for older adults~\cite{harrington_etal_DesigningHealthFitness_2018}. Similarly, another study examined the effectiveness of using audio-recorded educational materials framed in local historical stories to support computer skills training in rural areas~\cite{jonas_hanrahan_DigitalStorytellingDeveloping_2022}. Aligned with these studies, our findings indicate the challenges caused by the low digital-literacy in the farming setting, such as the difficulties in using Excel to keep track of crop information, indicating the need to design simple and easy-to-use digital tools to support small-scale farming communities in tracking information.

Recent work in rural computing found that older farmers in the Global North employ similar alternative data tracking techniques using physical formats (e.g., writing on printed templates on paper)~\cite{friedman_etal_UnderstandingFarmersData_2024}. These studies suggested designing applications for mobile phones that farmers are familiar with and frequently use~\cite{friedman_etal_UnderstandingFarmersData_2024}, along with wearable technologies that require minimal digital interaction~\cite{chaudhary_etal_NormalizingGritFutility_2025}, to support farmers' data tracking practices. These approaches can be extended to other farming communities, including Hmong farmers. 

Drawing on our findings and insights from previous studies, we recommend that information tracking tools should adopt farmers’ preferred workaround practices. These tools could leverage simple mobile technologies such as voice-based reporting, photo capture, and map tagging to support community-wide information sharing without requiring advanced digital literacy. Mobile applications or lightweight smart systems can also help farmers record farming schedules and activities using voice input or basic sketch recognition. Farmers could log crop types, planned planting dates, and associated field areas through verbal commands or uploaded sketch photos, and the system could provide timely reminders to reduce missed plantings and ease the burden of managing complex schedules.

In addition, given the strong bonding capital in the community, information tracking tools designed for this group could also support collective information sharing. Farmers often learn about emerging issues such as pest outbreaks or crop diseases through conversations with peers or by observing signs during routine fieldwork. Digital tools, such as a mobile application, could enable them to quickly document and share observations with others in their network. These tools might include features like voice messaging, photo uploads with brief notes, or map-based tags, making them accessible to farmers with varying levels of literacy. For example, farmers can speak into the app to share unusual spots on leaves with community members, while others with relevant insights can provide suggestions in suitable formats and share them with fellow farmers to prevent severe consequences. Similarly, such tools could help farmers track crop rotation patterns across shared plots or fields. In addition to tracking issues on the farm, these tools can assist farmers in sharing information about the types and quantities of crops they are growing, as well as identifying crops with high yields. When farmers observe that a particular crop is being overproduced, they can adjust their planting schedules and shift to other crops. This strategy helps prevent negative impacts on market prices and fosters mutual support within the community. Over time, this shared tracking and response system can contribute to collective knowledge building, where solutions are co-created and refined by members of the community.

\section{Limitations and Future Work}
Our study explores the practices and experiences of one small farming community, Hmong farmers, and therefore presents several limitations. Our results reflect the specific characteristics of this community and the unique forms of social capital that shape their work. This focus leaves other culturally distinct small-scale farming communities underexplored. Additionally, despite our emphasis on rapport-building, researchers’ own positionalities and biases may have influenced how we understood, framed, and interpreted Hmong farmers’ routines and practices.
Future work will build on this foundation by examining other small-scale farming communities whose histories, organizational structures, and social networks differ from those of the Hmong farmers in this study. Such work can reveal additional forms of bonding, bridging, and linking social capital and further illustrate how these relationships shape farmers’ strategies, constraints, and technology needs.

\section{Conclusion}
In this study, we interviewed 11 Hmong farmers from local community-based small farms and identified four key themes, including farmers’ workaround strategies to address challenges related to information tracking and the lack of adequate farming tools. Drawing from our findings, we discussed three types of social capital (bonding, bridging, and linking) and highlighted how these are uniquely fostered in small farming communities through intergenerational knowledge sharing, mutual support networks, and strong local relationships. These insights encourage HCI researchers to consider the specific dynamics of social capital in small farming communities when designing technologies and to account for the practical needs shaped by their farming practices. This study highlights the strengths and opportunities present in small farming communities, offering guidance for designers and researchers seeking to develop contextually grounded and sustainable technology-mediated interventions.

\begin{acks}
The authors thank the Hmong American Farmers Association (HAFA), The Good Acre, and the participating farmers for sharing their experiences with us. This work was supported by the University of Minnesota Imagine Faculty Research Grant and the Minnesota Agricultural Experiment Station (MAES).
\end{acks}

\bibliographystyle{ACM-Reference-Format}
\bibliography{main}

%%\appendix

\end{document}